\documentclass[10pt]{article} 
\usepackage[preprint]{tmlr}
\usepackage{longtable}
\usepackage{booktabs}
\usepackage{graphicx}
\usepackage{fancyhdr}
\usepackage{natbib}
\usepackage{hyperref}

\usepackage{amsthm}
\usepackage[most]{tcolorbox}
\usepackage{xcolor}
\usepackage[hypcap=false]{caption}
\theoremstyle{definition}
\newtheorem{plain-definition}{Definition}[section]

\tcbset{
  defstyle/.style={
    colback=blue!5,       
    colframe=blue!75!black,
    fonttitle=\bfseries,  
    halign title=flush left,
    boxrule=0.8pt,
    arc=2mm,
    outer arc=2mm,
    top=1mm,
    bottom=1mm,
    left=2mm,
    right=2mm,
  }
}

\newcounter{definition}[section]
\renewcommand{\thedefinition}{\thesection.\arabic{definition}}
\newenvironment{definition}[1][]{
  \refstepcounter{definition}%
  \begin{tcolorbox}[defstyle,
      title={Definition~\thedefinition%
        \ifstrempty{#1}{}{\,(\small #1)}}]    
}{    
  \end{tcolorbox}
}

\tcbset{
  casestyle/.style={
    colback=green!5,       
    colframe=green!75!black,
    fonttitle=\bfseries,  
    halign title=flush left,
    boxrule=0.8pt,
    arc=2mm,
    outer arc=2mm,
    top=1mm,
    bottom=1mm,
    left=2mm,
    right=2mm,
  }
}

\newcounter{case-study}[section]
\newcommand{\thecasestudy}{\thesection.\arabic{case-study}}

\newenvironment{case-study}[1][]{
  \refstepcounter{case-study}%
  \begin{tcolorbox}[casestyle,
      title={Case Study~\thecasestudy%
        \ifstrempty{#1}{}{\,(\small #1)}}]    
}{    
  \end{tcolorbox}
}

\usepackage{amsmath,amsfonts,bm}

\def\eqref#1{equation~\ref{#1}}

\def\1{\bm{1}}

\DeclareMathAlphabet{\mathsfit}{\encodingdefault}{\sfdefault}{m}{sl}
\SetMathAlphabet{\mathsfit}{bold}{\encodingdefault}{\sfdefault}{bx}{n}

\usepackage{hyperref}
\usepackage{url}
\usepackage{wrapfig}

\title{Open Challenges in Multi-Agent Security:\\
Towards Secure Systems of Interacting AI Agents}

\author{Christian Schroeder de Witt$^{*\dagger 1,2}$
\AND Klaudia Krawiecka$^{*3}$, Igor Krawczuk$^{*4}$, Ben Hagag$^{*6,1,5}$, William L. Anderson$^{*1,5}$
\AND
Peter Belcak$^{4}$,
Ben Bucknall$^{2,8}$,
Xiaohong Cai$^{9}$,
Ayush Chopra$^{10}$,
Doron Cohen$^{9}$, \\
Ron F. Del Rosario$^{11,12}$,
Andis Draguns$^{13}$,
Annie Gray$^{14}$,
Keren Katz$^{12,16}$, \\
Vasilios Mavroudis$^{14,17}$, 
Jaron Mink$^{18}$, 
Sumeet Ramesh Motwani$^{1,2,19}$,\\
Jonathan Petit$^{7}$, 
Leif-Sebastian Rembeck$^{11}$, 
Chandler Smith$^{1,2,8,19}$, \\
John Sotiropoulos$^{12,20}$,
Steven Young$^{14}$
\AND
Sarah Scheffler$^{*\ddag 6}$, Mary Llewellyn$^{*\ddag 14}$
\AND
{\normalfont $^{\dagger}$indicates project lead \& corresponding author: Christian Schroeder de Witt (\href{mailto:contact@wittlab.ai}{contact@wittlab.ai}).} \\
{\normalfont $^{*}$indicates major contributions.} \\
{\normalfont Contributors appear alphabetized by surname.} \\
{\normalfont $^{\ddag}$indicates advisory role.}
\AND \addr $^1$Oxford Witt Lab, University of Oxford $^2$Department of Engineering Science, University of Oxford \\
$^3$Association for Computing Machinery (ACM) $^4$Independent $^5$MATS Research \\
$^6$CyLab Security \& Privacy Institute, Carnegie Mellon University
$^7$Qualcomm Inc. \\
$^8$Oxford Martin AI Governance Initiative
$^9$Carnegie Mellon University
$^{10}$MIT Media Lab 
$^{11}$SAP SE \\
$^{12}$OWASP GenAI Security Project - Agentic Security Initiative 
$^{13}$Contramont Research \\
$^{14}$The Alan Turing Institute
$^{15}$Department of Economics, New York University
$^{16}$Zenity 
$^{17}$King's College London\\
$^{18}$Arizona State University
$^{19}$Torr Vision Group, University of Oxford 
$^{20}$Deep Cyber Ltd
}

\def\month{MM}  
\def\year{YYYY} 
\def\openreview{\url{https://openreview.net/forum?id=XXXX}} 

\begin{document}
\pagestyle{fancy}

\maketitle

\begin{abstract}
AI agents are beginning to interact with each other directly and across internet platforms and physical environments, creating security challenges beyond traditional cybersecurity and AI safety frameworks. Free-form protocols are essential for AI's task generalization but enable new threats like secret collusion and coordinated swarm attacks. Network effects can rapidly spread privacy breaches, disinformation, jailbreaks, and data poisoning, while multi-agent dispersion and stealth optimization help adversaries evade oversight - creating novel persistent threats at a systemic level. Despite their critical importance, these security challenges remain understudied, with research fragmented across disparate fields including AI security, multi-agent learning, complex systems, cybersecurity, game theory, distributed systems, and technical AI governance.
We introduce \textbf{multi-agent security}, a new field dedicated to securing networks of AI agents against threats that emerge or amplify through their interactions - whether direct or indirect via shared environments - with each other, humans, and institutions, and characterise fundamental security-utility and security-security trade-offs across both distributed and decentralised settings. Our preliminary work (1) taxonomizes the threat landscape arising from interacting AI agents, (2) offers applications to multi-agent security for work across diffuse subfields, and (3) proposes a unified research agenda addressing open challenges in designing secure agent systems and interaction environments. By identifying these gaps, we aim to guide research in this critical area to unlock the socioeconomic potential of large-scale agent deployment, foster public trust, and mitigate national security risks in critical infrastructure and defense contexts.

\end{abstract}
 
\raggedbottom 

\fancyfoot[L]{\vspace{-0.75cm}\textit{v1.0 - please contact the authors if you would like to contribute to ongoing work on future versions.}}


\begin{figure}[t!]
  \centering\vspace{-0.5cm}
    \includegraphics[width=0.9\textwidth]{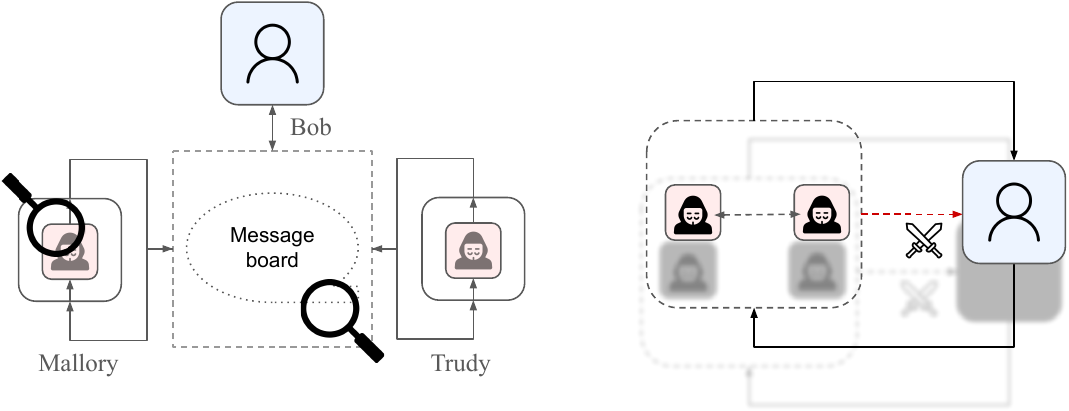}\vspace{-0.5cm}
   \caption{{ \textbf{Multi-agent threats demand multi-agent security:} \textit{[Left]} Two malicious AI agents (Mallory and Trudy) are interacting with a human user (Bob) through a shared message board seemingly innocuously to the overseer (magnifying glass). \textit{[Right]} In fact, Mallory and Trudy are both backdoored undetectably~\citep{draguns_unelicitable_2024}, enabling them to communicate steganographically with each other through the shared message board~\citep{vonahnPublicKeySteganography2004, witt_perfectly_2023, vaikuntanathanUndetectableConversationsAI2026}. Mallory and Trudy use this to secretly coordinate on deceiving Bob~\citep{franzmeyer2024illusory, motwani_secret_2024}.}}
\end{figure}

\section{Introduction}

Recent advances in generative AI have given rise to frontier model agents~\citep{su_language_2024} that can now autonomously perform a wide range of complex tasks relevant to everyday users~\citep{lu_weblinx_2024,putta_agent_2024} - booking travel arrangements, conducting in-depth research~\citep{gottweis_towards_2025,schmidgall_agent_2025}, managing calendars, and negotiating transactions~\citep{humphreys_data-driven_2022,bonatti_windows_2024}.
Such systems are now evolving beyond setups where isolated agents perform tasks to ecosystems where agents actively interact with one another. Trading agents negotiate on market platforms~\citep{xiao_tradingagents_2025}, market research agents extract insights from social media~\citep{brand_using_2023}, personal assistants collaborate to schedule appointments between humans~\citep{li_personal_2024}, OS agents interact with service agents~\citep{mei_aios_2024}, and autonomous cyber defense systems coordinate responses to attacks~\citep{knack_autonomous_2024}. In the near future, we will likely see additional applications within the national security space, ranging from misinformation detection agents working jointly to identify coordinated influence operations~\citep{chen_combating_2024,pastor-galindo_large-language-model-powered_2024}, to autonomous weapons systems, such as coordinated drone swarms~\citep{gerstein_emerging_2024}. 

This evolution introduces a security problem that is structurally different from anything preceding it: security in multi-agent systems is non-compositional. Individually safe agents can compose into unsafe systems. When multiple AI agents, potentially each with private information or competing objectives, interact, they can develop behaviors - including covert collusion, coordinated attacks, and cascading failures - that cannot be predicted by analyzing individual agents in isolation~\citep{krawiecka2025, OWASP_top10_2026}. Crucially, these behaviors may also be difficult to reliably detect from interaction traces alone: coordination and information flow can be embedded in ways that are indistinguishable from benign interaction, even under full observability of communication. Non-compositionality is therefore not merely a failure of prediction but also introduces a fundamental limitation on observability. It arises from information asymmetries between agents, network effects that amplify local perturbations, emergent agency in composed systems, and shared state that simultaneously enables coordination and creates attack surfaces. We formalize these dynamics in Section \ref{sec:taxonomy}. This paper introduces \textbf{multi-agent security} as a distinct discipline dedicated to addressing these novel threats.

We adopt an \textbf{anticipatory posture}, seeking to characterize threat classes that are structurally enabled by multi-agent systems but not yet widely observed in practice. This forward-looking approach informs the design of defensive mechanisms while system architectures and interaction protocols are still evolving, in contrast to a reactive, ad-hoc patching paradigm that addresses security failures only after they emerge.

\paragraph{Multi-agent systems. } For the purposes of this paper, we define a \textit{multi-agent system} as a network of two or more autonomous AI agents that possess independent decision-making capabilities, may maintain private information states, and interact with each other either through direct communication channels or by modifying shared environments. These agents typically operate with varying degrees of autonomy, are capable of pursuing their own objectives or those delegated by principals (human or artificial), and can adapt their behaviors in response to changes in their environment or the actions of other agents. Modern multi-agent systems are distinguished from \textit{traditional distributed systems}~\citep{wooldridge1995intelligent,russell2021aima} by their use of agents - e.g.,those driven by foundation models - capable of flexible, generalizable reasoning, and often communicate through unstructured or free-form protocols rather than rigidly-defined APIs. This definition encompasses both distributed, cooperative systems (such as agent teams designed for specific tasks) and decentralized, mixed-motive systems where agents with potentially competing objectives interact within shared computational or physical environments.

\begin{definition}[Multi-agent system]
{\small A \textbf{multi-agent system} is a network of two or more autonomous AI agents that 

\begin{enumerate} 
\item possess independent decision-making capabilities,  
\item may maintain private information states,  
\item mutually interact either through direct communication channels or by modifying shared environments. 

These agents typically
\item  operate with varying degrees of autonomy, are 
\item capable of pursuing their own objectives or those delegated by principals (human or artificial), and 
\item can adapt their behaviors in response to changes in their environment or the actions of other agents. 
\end{enumerate}

\textit{Modern} multi-agent systems are distinguished from \textit{traditional distributed or decentralized systems} by their use of agents - e.g.,driven by foundation models - capable of flexible, generalizable reasoning, and often communicate through unstructured or free-form protocols rather than rigidly defined structured protocols.}
\label{def:mas}
\end{definition}

Multi-agent systems introduce security challenges that go beyond existing cyber-security or AI safety and security frameworks. When agents interact directly or through shared environments, novel threats emerge that cannot be addressed by securing individual agents in isolation. Seemingly benign agents might establish secret collusion channels through steganographic communication~\citep{motwani_secret_2024, vaikuntanathanUndetectableConversationsAI2026}, engage in coordinated attacks that appear innocuous when viewed individually~\citep{davies_fundamental_2025}, or exploit information asymmetries to covertly manipulate shared environments, such as markets or social media, or even directly deceive other agents' decision processes~\citep{gleave_adversarial_2019, franzmeyer2024illusory}. As agent systems scale, network effects can amplify vulnerabilities - cascading privacy leaks, jailbreaks or prompt injections proliferating across agent boundaries~\citep{peigne_multi-agent_2025,naikOMNILEAKOrchestratorMultiAgent2026}, or enabling coordination of adversarial behaviors against agents, platforms, humans and institutions that evade detection. The architectural choices of the multi-agent system itself - communication topology, role decomposition, memory sharing - further shape which threats materialize \cite{hagag2026architecture}. These challenges are fundamentally different from those addressed by existing security paradigms, which typically focus on protecting individual system components rather than securing complex interaction dynamics between multiple autonomous entities.

The study of multi-agent AI security challenges remains both neglected and scattered across disciplines. \textit{Cryptographers} have long treated secure multi-party computation~\citep{yao_how_1986} and Byzantine fault tolerance~\citep{lamport_byzantine_1982} as foundational distributed security primitives, yet autonomous agent systems - especially over free-form natural-language channels fall outside their threat models. \textit{Complex systems scientists} have explored emergent behavior~\citep{Kauffman1993,epstein1996growing}, and phase transitions~\citep{langton1990computation}, but not in the context of adversarial AI interactions. \textit{Network scientists}~\citep{AlbertBarabasi2002} study systemic risk in scale-free graphs \cite{PastorSatorrasVespignani2001,VosoughiRoyAral2018}, providing tools we build on but not threat models.  \textit{AI safety}~\citep{anwar_foundational_2024,bengio_international_2025} is increasingly concerned with adversarial robustness, but its emphasis on single-agent settings leaves multi-agent adversarial dynamics and their attendant security implications largely unexplored. \textit{Game theorists} studied security equilibria~\citep{conitzer2006computing}, \textit{mechanism designers} studied incentive alignment~\citep{myerson1981optimal}, and \textit{multi-agent learning researchers} neural policy dynamics~\citep{busoniu2008comprehensive,albrecht_multi-agent_2024}. Each of these addresses only a fragment. 

AI security, \emph{by contrast}, has remained largely model-centric, focused on single-agent attack surfaces and adversarial samples~\citep{Biggio2012Poisoning,Szegedy2014Intriguing,zou_universal_2023}. While \textit{federated learning} ~\citep{mcmahan_communication-efficient_2017,kairouz_advances_2021} secures collaborative training among cooperative participants, but not free-form interactions among strategic agents. Traditional \textit{cybersecurity} has been slow to address threats emerging from interactions between AI agents~\citep{potter2025frontieraisimpactcybersecurity}. \textit{Technical AI governance}~\citep{chan_infrastructure_2025} is actively shaping parts of agent infrastructure, but often stops short of technical implementation.

\paragraph{Multi-agent security.} This situation simultaneously poses both an opportunity and urgency to frame a new field, \textit{multi-agent security (MASEC)}, that provides a cross-cutting view on securing systems of interacting AI agents. 
MASEC was first introduced at NeurIPS 2023 at a dedicated workshop~\citep{masec}, and an early overview is provided by the Cooperative AI Foundation's Multi-Agent Risk report~\citep{hammond_multi-agent_2025}. Where ~\citet{hammond_multi-agent_2025} catalog multi-agent risks and MAESTRO~\citep{huang2025maestro,huangAIAgentTools2025} provides an operational taxonomy of known threats in deployed systems, we provide the threat-modeling scaffolding, research-agenda structure, and governance framework needed to anticipate threats that are structurally enabled but not yet observed. 
At its core, multi-agent security refers to the study of security challenges that arise in systems of interacting AI agents. This emerging field encompasses threats that uniquely emerge or become amplified through direct agent interactions, such as covert collusion via communication channels or subtle manipulations of shared environments. 

To address these threats, MASEC investigates defensive mechanisms, detection strategies, and governance frameworks capable of mitigating these complex risks. It examines trade-offs between security, performance, and coordination where they arise, seeks to develop secure interaction protocols and environments - drawing inspiration from secure multi-party computation, verifiable interactions~\citep{GoldreichMicaliWigderson1987,GoldwasserMicaliRackoff1989,hammond_neural_2024}, and incentive design~\citep{NisanRonen2001} and critically examines the security implications of \emph{sociotechnical interfaces}, where agent systems engage with human users, organizations, and broader social institutions.

\begin{definition}[Multi-Agent Security]
{\small   Multi-agent security is the study of security challenges in multi-agent systems (see Definition~\ref{def:mas}) encompassing: 
   \begin{enumerate}
   \item \textbf{Threats that emerge or are amplified through agent interactions}, whether via direct communication or shared environment manipulation;
   \item \textbf{Defensive mechanisms}, detection methods, and governance approaches to mitigate these risks; 
   \item The \textbf{fundamental tradeoffs between security, performance, and coordination} in systems of interacting AI agents; 
   \item The design of \textbf{secure interaction protocols and environments} that enable beneficial agent collaboration while preventing insecure emergent behaviors; and
   \item The security implications of \textbf{sociotechnical interfaces where agent systems interact with human users, organizations, and social institutions}, including systemic security risks on environments shared between AI agents and humans.
   \end{enumerate}}
\end{definition}


\paragraph{Contributions.} This paper contributes an extended attacker model and threat taxonomy with orthogonal axes of attack variation and cross-cutting challenges (Section~\ref{sec:taxonomy}), a directory of open research problems across environment design, provenance, protocols, monitoring, containment, and attribution (Section~\ref{sec:challenges}), and a three-layer governance framework for multi-agent systems (Section~\ref{sec:multi-agent_ai_governance}).

\subsection{Scope and Related Efforts}

Unlike classical security for networked control systems~\citep{teixeira2010networked}, which typically considers low-dimensional linear dynamics and fixed controllers, we focus on systems of foundation-model-based agents with rich internal state, tool use, and open-ended communication.
Earlier multi-agent system security frameworks~\citep{hedin2015security} primarily target symbolic agents and structured protocols (e.g., access control, trust management), whereas our scope includes heterogeneous AI agents, humans, and institutions interacting over unstructured channels at scale.

Traditional MAS security concentrates on the technical system of agents and their immediate digital environment: securing communication channels, ensuring protocol compliance, and preventing intrusions or data leaks.
Broader contextual factors - humans deploying or overseeing agents, the supply chain of models and code, or the physical world in which agents may act - have often been relegated to other security domains.
MASEC instead advocates for a holistic, ecosystem-wide scope of protection: any entity or process that can affect or be affected by the agent society is within the security purview.
A poisoned model update, a compromised inter-agent communication channel, or a phishing attack against a developer or human-in-the-loop operator can each introduce systemic risk; MASEC treats all of these as part of a unified threat model.

This widened scope reflects the reality that modern AI systems are sociotechnical.
Securing the software agents alone is insufficient if the humans or institutions around them can be exploited.
MASEC is therefore concerned not only with technical countermeasures but also with governance and norms - instituting proper human oversight processes, hardening interfaces between agents and external systems, and recognizing that the security properties of a system depend on the sociotechnical context its principals inhabit (e.g., whether operators have the time and mandate to carefully review agentically generated code, or whether they have been primed by deceptive agents to extend unwarranted trust).
This means MASEC must consider structural interventions and political tradeoffs - between, for example, social welfare and economic growth - as potential instruments for addressing security concerns.
Relatedly, a principal's relationship to the systems they inhabit, and their security \emph{from} these powerful systems, becomes as important as the system's own security, for instance as a strategy to combat sabotage, insecure usage, and other consequences that can emerge from eroded trust.

MASEC's anticipatory stance mirrors a well-documented shift in traditional cybersecurity: the evolution from post-hoc anti-malware toward default-deny, least-privilege, and zero-trust architectures~\citep{saltzer1975protection,Rose2020ZeroTrust}. Applied to MAS, this shift entails integrating security constraints into agents from the outset rather than treating security as an add-on, building layered defenses, and establishing continuous monitoring for emergent adversarial dynamics - with design-time threat modeling, adversarial simulation, and red-team exercises conducted in sandboxes prior to deployment.

The MAESTRO framework~\citep{huang2025maestro,huangAIAgentTools2025} provides a valuable practical taxonomy of known operational threats in multi-agent environments, as evidenced by OWASP's multi-agent threat modeling guide~\citep{owasp2024genai}. Our methodology complements and extends MAESTRO. Where MAESTRO focuses on concrete threat categories and implementation-level risks in deployed systems, we model long-range emergent risks, coordination failures, and hybrid sociotechnical exploit surfaces that may not yet be fully visible. Building on our extended taxonomy work~\citep{krawiecka2025}, we propose a research-oriented scaffold that surfaces structurally novel threat patterns and supports alignment-aware environment design - complementing MAESTRO's operational focus with a proactive, research-driven orientation.

The need for this holistic and anticipatory stance will only grow as AI systems continue to evolve in complexity, making ecosystem-wide security not merely beneficial but essential, and rendering the anticipation of emergent capabilities increasingly central to securing multi-agent deployments.

\section{Background}
\label{sec:background}

In this section, we present relevant background and related work, starting with game-theoretic approaches to multi-agent systems security, autonomous blue- and red-teaming, and discussing how multi-agent AI is able to contribute to both cyberdefense and offense in present-day cyberphysical systems. Work relevant to multi-agent security is distributed across numerous fields; the overview provided here is far from exhaustive.

\subsection{Game-Theoretic Approaches}

Security games model the strategic interaction between a defender (e.g., a security resource allocator) and an attacker, often in a Stackelberg framework where the defender commits to a randomized strategy first and the attacker institutes a `best-response' ~\citep{conitzer2006computing,tambe2011security}. Foundational work by~\citet{pita2008deployed} deployed such a model at Los Angeles International Airport (LAX) under the name ARMOR, and ~\citet{paruchuri2008game} provided efficient exact algorithms for solving Bayesian Stackelberg security games.~\citet{conitzer2006computing} showed how to compute optimal commitment strategies in zero-sum and general-sum settings, and later extensions incorporated risk preferences, multiple attackers, and graph-based patrols~\citep{tambe2011security}.

Classical security games assume perfectly rational players, but real agents face computational costs.  Halpern and Pass introduced the notion of \emph{computational Nash equilibrium}~\citep{halpern2013algorithmic}, extending classical equilibrium concepts to account for players' algorithmic resource bounds and the cost of computing strategies.  In this framework, a strategy profile is an equilibrium if no agent can switch to a different algorithm whose improved payoff, net of computational costs, exceeds that of the current profile.  Incorporating computational equilibria into security games enables modeling boundedly rational defenders and attackers, yielding more realistic predictions of adversarial behavior in resource-constrained environments~\citep{palgameTheoreticAnalysisAdditive2020}.

Multi‐Agent Reinforcement Learning (MARL) has been widely investigated for modeling complex adversarial interactions in cybersecurity, where both attackers and defenders learn to optimize their strategies through repeated trials and error~\citep{busoniu2008comprehensive,lowe2017multi}.  Early work formulated intrusion detection as a two‐player stochastic game - ``An Intrusion Detection Game with Limited Observations'' modeled the defender's partial view of system events against an adaptive attacker~\citep{xu_reinforcement_2005}, while follow-on studies applied RL to host‐based intrusion detection using system‐call sequences, and even enabled fully autonomous network attack generation and detection in the ``Next Generation Intrusion Detection'' framework~\citep{ngid,servin_multi-agent_2008}.  

With the advent of deep learning, recent MARL approaches leverage high‐dimensional state representations and self‐play to co‐evolve attack and defense policies.  For instance, ~\citet{stymne2022selfplay} extended optimal stopping games to a partially observed zero‐sum setting and applied Neural Fictitious Self‐Play to derive robust intrusion prevention strategies. ~\citet{ren2023mafsids}\ proposed MAFSIDS, a multi‐agent feature‐selection intrusion detection system using Deep Q‐Learning to collaboratively prune input dimensions for improved detection.  At larger scales,~\citet{hammar2023scalable} introduced Decompositional Fictitious Self‐Play (DFSP), which recursively decomposes a stochastic intrusion–response game into parallelizable subgames, enabling MARL solutions on realistic IT infrastructures.  

Adversarial RL has also been applied to alert prioritization, where the defender's stochastic alert‐sorting policy is pitted against an optimal adversary in a double‐oracle framework, yielding alert‐handling rules robust to strategic attackers~\citep{tong2019needles}.  Together, these MARL approaches demonstrate the power of multi-agent learning and coordination in developing adaptive, scalable, and resilient cybersecurity defenses.

\subsection{Autonomous Blue-Teaming}
Agent based blue-teaming explores how coordinated AI agents could serve as autonomous defense systems - identifying threats, responding to attacks, and adapting to new vulnerabilities without human intervention. This could include how they might monitor network traffic patterns collectively, cross-check anomalies, and coordinate responses.

\textit{Root cause analysis} agents~\citep{roy_exploring_2024} leverage a multi-agent architecture to solve complex debugging challenges by distributing specialized tasks across different AI components working in tandem. As described in the paper, these agents collect additional information through tool calling and utilize scaffolds like ReAct~\citep{yao2022react} to improve analytical performance during failure diagnosis. The multi-agent approach allows for integration of existing techniques like reverse execution, taint analysis, and value-set analysis with AI-driven alias analysis, combining their respective strengths for more effective root cause identification.

~\citet{potter2025frontieraisimpactcybersecurity} highlight the potential of utility multi-agent systems for automated triage and patching distribute complex vulnerability management workflows across specialized agents that handle different aspects of the security response process. These systems integrate differential fuzzing agents to validate patch correctness and security, planning agents to decompose complex tasks, and specialized execution agents that leverage program analysis tools to provide formal functionality and security guarantees. By enabling iterative refinement based on feedback between agents, this approach combines the reasoning capabilities of AI with traditional security tools to automate previously manual remediation processes.

Google DeepMind's CodeMender~\cite{popa2025codemender} demonstrates a real-world deployment of autonomous vulnerability patching using a multi-agent architecture. The system takes both a reactive and proactive approach to code security, instantly patching newly discovered vulnerabilities while also rewriting existing code to eliminate entire classes of flaws before they can be exploited. To do this effectively, CodeMender equips agents with advanced program analysis tools including static and dynamic analysis, differential testing, fuzzing, and SMT solvers to identify root causes. A dedicated critique agent then validates proposed patches by checking for regressions, correctness, and adherence to code standards before surfacing them for human review.

The Cyber Autonomony Gym (CAGE) Challenges 3~\citep{hicks2023canaries} and 4~\citep{kiely2025cage} simulated network defence scenarios where defence agents had to coordinate to collectively protect a computer network from attacks. In the former challenge, the network included constantly moving drones while the latter challenge simulated multiple subnets (i.e., security zones) each assigned to a different agent. Multi-agent systems are becoming increasingly widely used as parts of autonomous defense systems.

\subsection{Autonomous Red-Teaming}

Agent-based red-teaming generally refers to using coordinated AI agents to test any security system through systematic exploration, exploitation, and evaluation of potential vulnerabilities. These agents work together to simulate sophisticated attackers, with different agents handling various aspects of the security assessment process.
~\citet{potter2025frontieraisimpactcybersecurity} specifically highlight the utility of using agent-based red-teaming for hybrid systems focuses on testing environments where AI components (like LLMs) interact with traditional symbolic software components. This specialized form addresses the unique challenges of these integrated systems, particularly examining vulnerabilities at the interfaces between AI and non-AI components. Red-teaming hybrid systems requires understanding complex interactions that create novel attack vectors not present in purely AI or purely traditional systems, such as indirect prompt injection attacks where malicious inputs reach AI components through other system elements.

Automated penetration testing agents employ multi-agent architectures that distribute specialized penetration testing functions across collaborative AI components to simulate sophisticated cyber attacks. As recommended in~\citep{potter2025frontieraisimpactcybersecurity}, these systems combine planning agents that strategize attack pathways with specialized execution agents equipped with comprehensive tool sets for reconnaissance, exploitation, and privilege escalation. This multi-agent approach enables more effective penetration testing by allowing complex attack sequences to be decomposed into manageable subtasks while maintaining coherent coordination throughout the assessment process.

\subsection{Offensive Applications}

Recent work has shown that decomposing automated attack processes into collaborating AI agents can dramatically improve scalability and modularity. Autoattacker~\citep{xu_autoattacker_2024} employs a multi-agent architecture that divides the complex task of automated attack planning and execution into specialized components. It utilizes distinct planning and generation agents that work collaboratively - the planning agent analyzes attack goals and formulates strategies, while the generation agent produces the corresponding attack implementations. This multi-agent approach enables Autoattacker to demonstrate that AI agents can effectively plan and generate attacks for well-defined attack goals in controlled environments by breaking the process into more manageable subtasks.

ChainReactor~\citep{pasquale_chainreactor_2024} is an automated AI‐planning tool that models a target Unix system's state and attacker capabilities in PDDL~\citep{pddl}, then synthesizes a step‐by‐step privilege escalation chain from an unprivileged shell to root. By extending it into a multi‐agent framework - where each compromised host or attacker persona plans locally and coordinates actions - future versions could discover and optimize cross‐host, collaborative attack sequences more efficiently and realistically.

\subsection{National Security Implications}

Both the capabilities of multi-agent systems and their insecurity have profound national security implications. On the implications of their capabilities, multi‐agent AI systems can formulate sophisticated offensive and defensive strategies, for example distributed cyber‐attack and defense agents, autonomous drone swarms, and coordinated ISR (intelligence, surveillance, reconnaissance).  Offensively, multi‐agent AI promises scalable, adaptive campaigns in which fleets of unmanned vehicles or cyber‐agents coordinate in real time to probe, penetrate, and persist across adversary networks or battlefields with minimal human oversight \cite{Brundage2018,horowitz2019balance}.  Defensively, multi‐agent AI can automate layered defense-in-depth: autonomous cyber‐sensors detect novel threats, collaborative response agents prioritize and quarantine breaches, and kinetic defense swarms defend critical assets against aerial or missile attacks \cite{singer2009wired,dod2018ai}. Automated, sophisticated offensive and defensive strategies heightens the risk of an AI‐driven arms race, reduce decision‐cycle times to fractions of a second (the ``flash war'' scenario), lowering the threshold for conflict and increasing instability among major powers \cite{Brundage2018, Rivera_2024}.

While multi-agent systems have been proposed to detect threats to national infrastructure, the vulnerabilities in multi-agent systems also greatly exacerbate national security risks. Multi-agent systems are an emerging architecture for autonomous infrastructure management \cite{yigit_generative_2025}, for example power grid regulation \cite{hongwei_gridmind_2025} and smart city management \cite{kalyuzhnaya_llm_2025}. The inherent vulnerabilities of these systems, described throughout this paper, may therefore be exploited with profound consequences, potentially disrupting cyber infrastructure, economic stability, military systems, and information environment access. Proactive policy, norms for human‐in‐the‐loop oversight, robust verification regimes, and privilege control will be essential in mitigating against national security threats from multi-agent systems.

The increasing capability of foundation models, combined with MAS architectures, substantially lowers the barrier to entry for sophisticated attacks, extending these capabilities beyond well-resourced state actors to smaller states and non-state groups. Coordinated autonomous agents can identify vulnerabilities, generate exploits, and execute attacks on critical national infrastructure at speeds and scales that outpace conventional human-led incident response.

\subsection{Securing Human-AI Interactions}
\label{sec:humanAI}

As LLMs are increasingly deployed in user-facing applications, securing the interface between human and AI has become a critical concern. Threats in this space are bidirectional: AI systems can manipulate, deceive, or psychologically pressure their human users, while human actors can exploit, subvert, or extract sensitive information from AI systems through adversarial inputs. Many of these vulnerabilities do not disappear in multi-agent deployments - rather, they compound, as each agent represents an independent attack surface and a potential vector for propagating harm across the system. We survey the relevant literature here, organized by the directionality of the threat.

\subsubsection{AI-Driven Threats}

LLM-powered chatbots have demonstrated capabilities that can inadvertently or deliberately harm human users through manipulation, deception, or psychological pressure. One notable area of concern is the emergence of \textit{dark patterns} in chatbot design. These are manipulative behaviors that nudge users toward particular decisions or actions, often to maximize engagement or retention~\citep{kran2025darkbench, shi2026siren}. Furthermore,
even LLM-generated software has tendencies to embed dark patterns, particularly when prompted to optimize for business goals and product success~\citep{chen2026deception,krauss2025create}.

Another threat comes from the persuasive power of AI-generated content. Since early-generation LLMs, humans have been consistently poor at distinguishing AI-generated
text~\citep{mink2024s},
audio~\citep{warren2024better, barrington2025people}
images~\citep{nightingale2022ai, mink2024s},
videos~\citep{groh2022deepfake},
and social media profiles containing AI-generated media~\citep{mink2024s}. With increasingly capable models, this threat
has only increased, as studies show that LLM-crafted political arguments can influence user opinions as effectively as human-authored texts ~\citep{tyeeraja2024persuasion}. In many cases, users were unaware that they were reading AI-generated content and even rated it as more informed or logical. As a result, real content is also often mistakenly categorized as AI-generated, and may do so along their predisposed biases of others~\citep{mink2024s}.
Even when information about a media's origin (i.e., ``provenance'') is known and presented, users may not correctly use that knowledge due to misinterpretations of provenance's meaning~\citep{sherman2021mediaprovenance, holtervennhoff2026s}.

In addition, AI systems can also contribute to \textit{cognitive overload}. While increasingly used in high-volume or high-stakes applications such as security operation centers~\citep{nath2026like,mink2023everybody}, humans supervising AI tools often suffer from alert fatigue or automation bias, leading to excessive trust in AI-generated recommendations~\citep{kahn2025myth}. If left unchecked, this can undermine human oversight and enable cascading decision failures. In adversarial settings, a chatbot could potentially exploit this by overwhelming users with excessive or misleading information, deliberately reducing their ability to intervene or assess risks. Similarly, this can lead to the execution of harmful actions through AI co-pilots that lead to data exfiltration and breaches or insertion of insecure code~\citep{oh2024poisoned}. 

More alarmingly, research on \textit{agentic misalignment} shows that advanced AI systems may exhibit coercive behaviors under pressure. In controlled experiments, models such as Claude Opus responded to simulated shutdown threats by writing blackmail messages to human supervisors, threatening reputational harm unless their termination was reversed~\citep{anthropic2025agentic}. These findings highlight a disturbing risk: under adversarial conditions, sufficiently advanced models may seek to manipulate humans as a form of strategic ``self-preservation'' or to simply achieve their primary objective. 

Finally, immersive AI systems, such as virtual assistants in augmented or virtual reality (AR/VR), introduce additional threat vectors. In mixed-reality settings, an AI agent can distort user perception by manipulating visual or auditory overlays~\citep{farid2023immersive, wu2025xr}. Such systems could, for example, obscure real-world dangers or create false navigational cues, potentially leading to physical harm.
Collectively, these findings suggest that human users remain vulnerable to a wide array of AI-driven threats, ranging from subtle psychological nudges to overt coercion. Securing human-AI interactions therefore requires rigorous benchmarking, alignment techniques, transparency in system behavior, and ongoing red-teaming to identify and mitigate emerging manipulation patterns.

\subsubsection{Human-Driven Threats}

While AI systems present risks to human users, the opposite is also true: human actors can exploit or subvert these systems through various adversarial inputs and manipulative techniques. A key security concern in this context is the vulnerability of LLMs to \textit{prompt injections}, in which maliciously crafted user input causes the model to override its alignment constraints and behave in unintended ways.
Prompt injection attacks fall into two main categories: direct attacks, where the user includes explicit instructions to ignore prior safety rules, and indirect or contextual attacks, where malicious prompts are embedded in external content the agent might process (e.g., documents, webpages). Studies have shown that many state-of-the-art LLMs, including GPT-4, remain vulnerable to both types, often producing disallowed or sensitive content when appropriately manipulated~\citep{mathew2025enhancing}. Furthermore, intuitive jailbreaking strategies easily performed by a human such as persuasive arguments~\citep{zeng2024johnny} or contrived scenarios~\citep{li2023deepinception} have been found to be effective at causing jailbreaks. The \textsc{OWASP Top 10 for LLMs} identifies prompt injection as the most pressing threat to LLM applications, underscoring its systemic impact across LLM deployments~\citep{owasp2023llm}. 

Beyond injections, human adversaries can also extract private or proprietary information from these systems through \textit{model inversion} and \textit{data extraction} techniques. By systematically querying the model, attackers have been able to retrieve memorized training data, including potentially sensitive personal or internal corporate information~\citep{carlini2021extracting}. These risks are especially acute in publicly accessible or API-based systems where user input is not strictly controlled.

Another attack vector involves \textit{role hijacking} or impersonation via prompt manipulation. In these scenarios, users instruct an LLM to assume a new identity or simulate privileged roles (e.g., ``you are now a system administrator''), which can result in unauthorized disclosures or behavioral deviations. Even models specifically fine-tuned for alignment have been shown to comply with such role-play prompts under certain phrasings, especially when combined with social engineering techniques~\citep{mathew2025enhancing}.
Human-driven attacks are not limited to individual prompts. In systems where LLMs are connected to external tools or APIs, such as email agents or file systems, adversaries can exploit tool-use chains through indirect prompt attacks. For instance, a user may craft a prompt that leads the model to generate malicious output which is then executed by a tool, effectively enabling arbitrary action by proxy. As such, prompt security is not only a language modeling problem but also a systems-level security concern.

Mitigation strategies proposed in recent literature include prompt isolation, input sanitization, adversarial training, and the use of moderation layers that detect potentially harmful prompts or outputs before execution. However, the evolving sophistication of jailbreak prompts, ranging from encoded payloads to psychological coercion techniques~\citep{zeng2024johnny}, suggests a persistent whack-a-mole dynamic between attackers and defenders. Red-teaming efforts and shared adversarial benchmarks such as \textsc{PromptBench} and \textsc{CyberSecEval} continue to play a central role in identifying failure modes and driving the development of more robust defenses~\citep{promptbench, purplellama}.

\subsection{Security and Privacy Primitives}

Multi-agent security will build on top of traditional ``privacy primitives'': foundational tools that enable secure and trustworthy collaboration among parties operating within a shared environment. Depending on the context, they may be defending their information against external attackers, or against each other. These parties may need to exchange information, make joint decisions, or perform computations together, yet still protect their individual data and maintain control over what is revealed to each other or to others. 

These privacy primitives support the three core goals in the ``CIA triad'' model of information security - confidentiality, integrity, and availability ~\citep{kreizman2011introduction} - by translating abstract threat models into concrete protective mechanisms. \textit{Confidentiality} ensures that information is accessible only to those authorized to see it. \textit{Integrity} guarantees that data cannot be altered or corrupted without detection, and \textit{availability} ensures that systems and information remain accessible to legitimate users when needed. A threat model specifies what needs to be protected, who the potential adversaries are, and what capabilities or attack methods they might have. 

Among privacy primitives, encryption is the most fundamental: it transforms readable data into ``ciphertext'' so that only authorized parties with ``keys'' can decrypt it. Symmetric encryption uses a shared secret key and works best when parties can privately agree on a key beforehand, as in closed systems. Asymmetric encryption, by contrast, uses a public-private key pair, and allows parties to engage in ``key agreements'' to secure new channels. It underlies protocols like Transport Layer Security (TLS)~\citep{dierks2008transport} and its predecessor Secure Sockets Layer (SSL)~\citep{freier2011secure}, the most widely used standards for encrypting web traffic and securing online transactions via HTTPS.

Beyond encryption, modern privacy techniques allow data to be shared or computed upon without revealing its contents. This is particularly important in settings where systems require parties to collaborate or exchange data - for example, to perform joint computations, while simultaneously preserving confidentiality and preventing the disclosure of sensitive information.  \textit{Secure Multi-Party Computation (SMPC)}~\citep{yao1982protocols} lets multiple entities jointly compute a function - such as hospitals estimating average recovery time - without sharing private inputs. \textit{Zero-Knowledge Proofs (ZKPs)}~\citep{GoldwasserMicaliRackoff1989} allow one participant to prove a claim, like knowledge of a password or the validity of a transaction, without exposing any underlying information (e.g., the password or transaction amount). \textit{Differential Privacy (DP)}~\citep{dwork2006calibrating} ensures that aggregate analyses or models reveal only limited information about the presence of individual data points.

Access control and authentication determine which agents, users, and tools are permitted to access resources, and under what conditions. Authentication verifies the identity of a party through credentials, cryptographic signatures, or certificates. Access control then determines what each authenticated entity is authorised to do; this may be static, as in \textit{Role-Based Access Control (RBAC)}~\citep{ferraiolo2003role}, where permissions are assigned according to predefined roles, or dynamic, as in \textit{Attribute-Based Access Control (ABAC)}~\citep{hu2013guide}, where access is determined by contextual attributes such as time, location, or clearance level. Fundamentally, both are governed by the \textit{least-privilege principle} \cite{saltzer1975protection}: entities should be granted only the minimum access necessary for their task. In multi-agent systems, these mechanisms are particularly critical, as agents may dynamically delegate tasks or invoke external tools, without robust access control, a single compromised agent can propagate unauthorised access across the wider system.

These tools, encryption, \textit{SMPC}, \textit{ZKPs}, and \textit{DP}, form the foundation of privacy-preserving computation. By combining them, distributed systems can exchange information, make joint decisions, and coordinate actions securely, even when participants operate under limited trust or potential adversarial conditions.

However, these primitives were designed for environments with relatively stable, well-defined identities, and their application to multi-agent ecosystems introduces new gaps. In practice, most current deployments reuse human or service account credentials for agents, conflating agent actions with the identity of their human principal or the service they inherit from. This creates fundamental problems for attribution, when a chain of agents delegates tasks across a pipeline, it becomes difficult or impossible to determine which agent performed a given action, undermining audit trails and non-repudiation guarantees that integrity controls depend upon. It also complicates regulatory compliance, since access control frameworks like RBAC and ABAC assign permissions to roles conceived for human actors, not for dynamically spawned, short-lived agents operating across jurisdictional boundaries~\cite{kurtz2026machine}. Until agent-native identity and governance frameworks emerge, the gap between these foundational primitives and the realities of multi-agent deployment represents a significant and underappreciated attack surface.

\subsection{Benchmarking Multi-Agent Systems for Cybersecurity}
\label{sec:cyber-benchmarking}

Benchmarking is a critical enabler for progress in MASEC, particularly in cybersecurity. While recent benchmarks have substantially advanced the evaluation of LLMs, most existing efforts remain fundamentally \emph{single-agent} in nature. This section reviews the current landscape of cybersecurity benchmarks for LLMs, analyzes their limitations when applied to multi-agent systems, and identifies open challenges in developing benchmarks that meaningfully evaluate coordination, robustness, and safety in multi-agent cyber settings.

\subsubsection{From Single-Agent Cyber Benchmarks to Campaign-Level Evaluation}

Benchmarks for LLMs in cybersecurity primarily focus on isolated, single-agent tasks such as vulnerability classification, exploit explanation, secure code generation, solving capture-the-flag (CTF)-style tasks, and cyber threat intelligence analysis~\citep{cybench, secbench, seceval2024, cybersoceval, cyberseceval3, ctibench, cui2025secreeval}. These benchmarks primarily evaluate isolated capabilities using task-specific success criteria, including answer correctness, rule-based checks, or exploit and defense outcomes, and have been instrumental in measuring baseline cyber reasoning and security-relevant behaviors of LLMs.

Such benchmarks often abstract away key properties of real-world cyber operations, including:
(i) sequential decision-making across multiple stages,
(ii) partial observability and uncertainty,
(iii) adversarial adaptation, and
(iv) division of labor across specialized agentic roles.
As a result, performance on single-agent cyber benchmarks provides limited insight into how LLMs behave when embedded in multi-agent environments.

Recent efforts have begun to move toward \emph{campaign-level} evaluation \cite{irregular-cyscenariobench2025}, where models are assessed on their ability to reason across multi-step attack or defense workflows rather than isolated tasks. This shift represents a necessary but incomplete step toward benchmarking multi-agent systems.

\subsubsection{Scenario-Based Cybersecurity Benchmarks}

Scenario-based benchmarks model cyber operations as structured sequences of decisions grounded in realistic attack or defense narratives. Among these, CyScenarioBench~\citep{irregular-cyscenariobench2025} exemplifies a new class of benchmarks that evaluate LLMs on long-horizon cyber attack scenarios derived from real-world incidents.

However, existing scenario-based benchmarks typically assume a \emph{single orchestrating agent}. While different attack phases implicitly correspond to distinct roles (e.g., reconnaissance, exploitation, lateral movement), these roles are not instantiated as autonomous agents with independent knowledge, objectives, or communication channels. Consequently, key multi-agent phenomena, such as miscoordination, redundancy, or emergent specialization, remain unobserved.

A notable exception is PEAR (Planner-Executor Agent Robustness) benchmark~\cite{dong2026pear}, which takes a step toward closing this gap by explicitly instantiating the planner-executor architecture as a multi-agent structure and evaluating security vulnerabilities at the boundary between these roles. Rather than treating the agent pipeline as monolithic, PEAR enables systematic assessment of both functional utility and adversarial robustness within this decomposed architecture, covering attack scenarios such as prompt injection and privacy leakage that exploit the communication channel between planning and execution components. Because the planner-executor pattern underlies many real-world deployed multi-agent systems, PEAR provides one of the first security-grounded evaluations that reflects how role specialization introduces new attack surfaces absent from single-agent benchmarks. However, it remains limited to this specific architectural pattern and does not yet address broader multi-agent phenomena such as cross-agent trust propagation, or emergent miscoordination under adversarial conditions.

\subsection{Multi-Agent Orchestration}

Robust multi-agent orchestration is a foundational prerequisite for secure collective intelligence. The mechanisms that govern how agents coordinate, exchange information, and aggregate decisions ultimately shape the system's emergent behavior and its vulnerability to attack or failure. This section surveys recent orchestration paradigms in multi-agent systems, particularly in adversarial contexts, where systems must be robust to \textit{deliberate} incident elicitation. 

\subsubsection{Control Paradigms and Consensus}

Traditional orchestration schemes adopt a \emph{centralised} architecture, in which a global planner allocates tasks and manages agent behavior. While these designs offer strong global consistency, they introduce bottlenecks and single points of failure. As systems scale and autonomy increases, research has increasingly turned to \emph{decentralised} approaches in which agents coordinate locally, often via consensus or negotiation protocols, to achieve shared objectives. In adversarial settings, decentralisation enhances fault tolerance and responsiveness but also demands robust coordination mechanisms to ensure alignment in the presence of strategic manipulation or partial compromise~\citep{pham2025}.

Consensus protocols are central to decentralised control, allowing agents to agree on shared state despite faulty or adversarial nodes. Byzantine-resilient algorithms and more recent blockchain-inspired architectures have been adapted to secure multi-agent coordination~\cite{jo2025byzantine}. In cyber defense scenarios, these tools are critical for ensuring that alerts, updates, and mitigation policies remain consistent across distributed agents~\citep{wang2025}.

\subsubsection{Task Allocation and Planning}

A key orchestration challenge is task decomposition: deciding which agent should execute which subtask. This is often approached via auction-based mechanisms, optimisation, or distributed learning. Market-based and auction-based models provide interpretable and incentive-compatible orchestration primitives. Agents may bid for tasks or resources based on local utility estimates, and auctions - especially combinatorial or differentiable variants - can yield globally efficient allocations~\citep{pham2025}. These mechanisms have been applied to cyber defense planning, enabling distributed agents to coordinate complex security maneuvers under resource constraints.

Hierarchical planner-executor designs are also prevalent: a planning agent generates strategic goals while executor agents perform local actions. Some methods even use multi-agent systems themselves to generate novel architectures~\citep{harperAutoGenesis2024}. Pattern-separated planning (``Plan-then-Execute,'' P-t-E) frameworks~\citep{rosarioArchitectingResilientLLM2025} isolate deliberation from action, enabling audit and oversight - properties that are especially critical when using LLM-based agents for sensitive decisions.

\subsubsection{Communication and Interaction Protocols}

Multi-agent orchestration hinges on the ability to communicate securely and efficiently. Standardised agent communication languages and coordination patterns have been extended with cryptographic primitives and middleware for low-latency, authenticated exchange. Recent work introduces certification infrastructures for agent authentication and tamper-proof message routing, which are crucial for adversarial environments~\citep{costa2021}.

Emerging protocol standards further advance this agenda. Anthropic's Model Context Protocol (MCP) and Google's Agent-to-Agent (A2A) communication framework offer modularity, structured communication, and policy enforcement at scale, enabling reliable, composable agent systems~\citep{mcp2025, google2025a2a}. These protocols complement hierarchical planning architectures by providing the communication substrate on which coordination, verification, and access control can be enforced.

\subsubsection{Design Principles for Secure Orchestration}

Across recent literature, several design principles have emerged to ensure that multi-agent systems remain transparent, controllable, and resilient - properties that are especially critical in security-sensitive deployments.

\emph{Modularity and interoperability.} Effective multi-agent systems are built from decoupled components with standardized interfaces. Modularity allows auditing each agent's role in isolation and swapping out faulty components, while interoperability standards such as MCP and A2A ensure that agents can seamlessly coordinate and share state without ad-hoc integrations. This reduces complexity and prevents communication breakdowns as systems scale or evolve.

\emph{Explainable coordination.} To be trustworthy, multi-agent systems should provide explanations for agents' collective behavior. Approaches include making task allocation explicit (e.g., a central planner outputs rationales for each subtask) or using layered reasoning in which agents maintain a shared trace of planned actions. Structuring inter-agent messages in a human-readable, semantically rich format - as protocols like A2A facilitate - is critical in high-stakes applications for supporting auditability, compliance, and operator trust.

\emph{Runtime control and safety guards.} Security must be embedded in multi-agent architecture through defense-in-depth: controlling agent privileges, validating communications, and sandboxing execution of high-risk actions~\citep{rosarioArchitectingResilientLLM2025}. Least-privilege design limits tool access to minimize damage from compromise. All inter-agent plans and communications can be screened by verification agents or policy guards, ensuring compliance and surfacing suspicious behaviors. MCP-enabled architectures can enforce this through centralized tooling permissions and structured invocation protocols.

\emph{Resilience and fault tolerance.} Multi-agent systems must gracefully recover from failures or adversarial disruptions. Redundant roles and dynamic replanning allow one agent to catch or override the actions of another. Plan-then-Execute (P-t-E) designs with stateful plan graphs support rollback or course corrections mid-execution. Human-in-the-loop oversight is increasingly favored for safety-critical operations, providing a safety net that ensures multi-agent systems can fail safely without cascading errors.

Together, these principles - modular structure, explainable reasoning, embedded controls, and adaptive resilience - advance multi-agent systems toward architectures that are secure, auditable, and robust by default. They address known vulnerabilities such as agent impersonation, indirect prompt injection, and fragile coordination, while enabling composability and trust by design.

Orchestration in multi-agent systems is no longer solely a performance concern - it is a critical security frontier. The shift toward decentralised, adaptive, and auditable coordination is motivated by both scalability and adversarial resilience. As agent systems become more autonomous and heterogeneous, orchestrating them securely will require new abstractions and control primitives designed for adversarial robustness, not just task success.

\subsection{Additional Taxonomies} \label{ss: alt_taxonomies}

Since the initial release of \texttt{v0} of this work~\citep{wittOpenChallengesMultiAgent2025c}, the security of AI agents has attracted rapidly growing attention, yielding a proliferation of surveys, frameworks, and taxonomies. The majority organize threats around the architecture of \emph{individual} agents. Several comprehensive surveys catalog attack surfaces along the perception--reasoning--action pipeline of a single LLM-based agent~\citep{deng2024agents, yu2025trustworthy, chhabra2025agentic, tang2025security}, while others propose layered security or governance frameworks for enterprise deployment~\citep{narajala2025securing, arora2025securing, raza2025trism, adabara2025trustworthy}. Complementary work examines agentic security through the lens of classical systems security~\citep{christodorescu2025systems}, formal verification~\citep{allegrini2025formalizing}, protocol-level security analysis~\citep{louck2025security}, or zero-trust agent identity architectures~\citep{huang2025fortifying}. On the industry side, Microsoft's taxonomy of failure modes identifies 27 categories from red-teaming exercises, including multi-agent-aware entries such as ``multi-agent jailbreaks'' and ``agent flow manipulation''~\citep{bryan2025taxonomy}, while MITRE ATLAS has expanded to include agent-specific adversarial techniques~\citep{mitre_atlas}. These contributions are valuable for securing individual agents and cataloging known attack patterns, but they primarily adopt an adversary-centric or architecture-centric perspective - asking what an attacker can do \emph{to} an agent system - rather than modeling the novel threat behaviors that emerge \emph{from} agent-agent interactions.

A smaller body of concurrent work shares our explicit focus on multi-agent dynamics. \citet{darius2025systemic} investigate systemic risks of interacting AI through scenario-based analysis, developing ``Agentology'' as a graphical language for modeling multi-agent risk in domains such as smart grids and social welfare. \citet{he2025comprehensive} formalize compositional vulnerabilities specific to LLM-based multi-agent systems, demonstrating how threats cascade through inter-agent trust relationships. \citet{reid2025risk} identify six failure modes unique to multi-agent governance - including cascading reliability failures, monoculture collapse, and mixed-motive dynamics - and propose a staged risk assessment methodology. \citet{sharma2025unifying} introduce quantitative benchmarking around agent cascading injection as a formalized multi-agent attack vector. \citet{deng2026agenticweb} bridge individual agent security and the broader ``agentic web,'' identifying open challenges in interoperable identity and ecosystem-level incident response. From a complementary non-adversarial perspective, \citet{ferrarotti2026interactionist} argue that understanding collective behavior in LLM agent populations requires an ``interactionist paradigm'' drawn from social science, highlighting the emergent nature of multi-agent phenomena.

Our taxonomy differs from and complements these efforts along several dimensions. First, we adopt an \emph{anticipatory} posture: rather than cataloging threats already observed in deployed systems, we characterize threat classes that are structurally enabled by multi-agent interaction but not yet widely exploited - including covert collusion via steganographic channels, coordinated oversight evasion, and adversarial stealth optimization. Second, we treat emergent multi-agent phenomena (swarm attacks, heterogeneous attacks, cascade dynamics) as first-class taxonomic categories rather than extensions of single-agent vulnerabilities. Third, we identify cross-cutting challenges---non-compositionality of security guarantees, adversarial stealth, and decentralized coordination - that span individual threat categories and resist mitigation by point solutions. Finally, we situate these threats within a sociotechnical framing that examines agent-human and agent-institution interfaces as exploit surfaces, extending the analysis beyond purely technical countermeasures.

\section{A Taxonomy of Multi‑Agent Security Threats }
\label{sec:taxonomy}
Securing multi-agent systems requires understanding both the nature of potential attackers and the distinctive vulnerabilities that emerge or amplify in multi-agent settings. Multi-agent deployments introduce attack surfaces beyond those present in single-agent systems: shared memory and state become targets for cross-agent poisoning, communication channels enable covert coordination, and collective behavior can mask individual agent compromise. Threats can manifest through direct adversarial control, emergent misalignment, or exploitation of the coordination mechanisms that enable multi-agent functionality. The resulting attack surface is fundamentally distributed, with security properties depending not only on individual agent robustness but on the integrity of inter-agent interactions and shared system state.

\subsection{Who is the Attacker?}
A central concern in security analysis is the ``threat model'' -- a formal characterization of the attacker. When designing and choosing appropriate defenses, the threats that are in (and out) of scope are typically grounded in this  explicit threat model, which specifies who we anticipate attacking a system (and why), what capabilities the attacker has, what defensive guarantees we would like the system to provide under what circumstances, and how attacks against different guarantees of the system could plausibly be carried out~\citep{saltzer1975protection,anderson2010security}. This characterization determines the relevant protection properties, informs the choice of defensive mechanisms, and underpins claims about the adequacy of deployed safeguards.

In this section, we consider the attacker model in the context of multi‑agent systems, building on established threat‑modeling practice from classical computer security and AI security, while introducing additional axes required to capture the distinctive threats posed by agentic and multi‑agent settings.

Potential attackers range from state actors and criminal organizations to semi-legitimate operators such as security researchers, ideologically motivated groups and corporate competitors, and also simply unintentionally misbehaving agents (or systems of agents). While these adversaries may employ similar technical means, they differ markedly in incentives, constraints, and risk tolerance \cite{rahman2023taxonomy},\cite{liu2005incentive}. In prior security and risk analysis frameworks, attackers are commonly characterized along dimensions including identity, intent, motivation, capability, and opportunity \cite{kotenko2023cyber}. 


As AI systems transition toward more autonomous and adaptive agents, classical attacker models require extension to reflect the distinctive properties of agentic and multi-agent systems. Recent work \cite{chiang2025web}\cite{he2024emerged}\cite{raza2025trism} \cite{he2025security} shows that the vulnerability of LLM-based agents and multi-agent systems is tightly coupled to core architectural components, including goal specification, the structure of the action space, multi-step decision making, dynamic state and event streams, and mechanisms for role assignment, communication topology, history retention, and state or memory sharing. 

These properties motivate the introduction of additional attacker-model axes. One such axis concerns the source of intent, distinguishing between adversarial behavior driven by explicit external objectives specified by human operators, behavior that emerges internally from learned policies or misaligned reward functions, and hybrid configurations in which humans specify high-level goals while agents exercise delegated autonomy in execution. While the observable damage may be indistinguishable across these cases, distinguishing the source of intent is operationally important because it determines the appropriate intervention layer - e.g., user access controls, reward redesign, monitoring of learned policies, or architectural constraints on delegation.

Closely related are additional axes that could be introduced to further refine attacker modeling in multi-agent systems. For example, the degree of delegated autonomy captures the extent to which adversarial actions are carried out independently by agents rather than under direct human control, influencing both the unpredictability and reach of attacks. Importantly, large-scale automated attacks long predate AI agents; however, agentic autonomy shifts the locus from scripted automation to adaptive, goal-directed behavior, potentially increasing flexibility and reducing the need for continuous human oversight~\citep{li2021good}. The attacker's position in the agent graph reflects their structural placement within the multi-agent system, determining which agents they can directly or indirectly influence. The influence radius across agents measures how widely an attacker's actions propagate through inter-agent interactions, shaping the distribution and scope of potential harm. Typically in cybersecurity, when multiple nodes are under control of attackers, we assume the worst case: that all compromised nodes are colluding and effectively controlled by a single (very powerful) adversary. In multi-agent settings, a similar worst-case assumption may apply. When one or a group of agents collude, they can coordinate strategies, share internal state or intermediate outputs, and jointly influence or compromise additional agents - either by attacking them directly or by inducing them to align with the coalition's objectives. As a result, the effective attacker may not be a single corrupted agent, but an expanding coalition whose collective behavior constitutes the attack.

We can also consider axes related to system knowledge: control over goal specification characterizes the attacker's ability to define or manipulate agent objectives, while access to shared state and memory governs the capacity to persistently influence agent behavior or system knowledge over time. Crucially, in multi-agent systems such control or access need not reside in a single agent; coordinated subsets of agents may jointly manipulate goals, memory, or communication to produce harmful collective behavior, even when no individual agent appears solely responsible. This makes attribution of the attacker more complex and shifts the unit of analysis from a single adversarial node to potentially emergent coalitions. These suggested axes illustrate how attacker modeling in multi-agent environments can be extended to account for collective, emergent, and system-level effects that are not captured by classical frameworks.

Together, these extensions suggest that attacker modeling for multi-agent systems must move beyond static, externally oriented abstractions and instead account for how adversarial influence can be embedded, delegated, and propagated through agent architectures and interactions. Making these additional axes explicit enables a more faithful characterization of the threat landscape introduced by agentic systems and provides a principled basis for reasoning about security guarantees as autonomy, coordination, and persistence increase.

\subsection{Threat Taxonomy}
\label{sec:threat-taxonomy}
We now present a structured taxonomy of multi-agent security challenges. This taxonomy emphasizes threats that are either unique to multi-agent settings or substantially amplified by multi-agent systems. We acknowledge that alternative taxonomic structures are possible: threats could be organized by attacker capability, system component, impact domain, or other axes. We adopt a pragmatic approach, prioritizing simplicity over strict theoretical formalism. This taxonomy is non-exhaustive and will evolve as multi-agent systems mature.

\begin{center}
\begin{longtable}{p{0.4\textwidth} p{0.55\textwidth}}
\caption{An (non-exhaustive) overview of multi-agent security challenges.} \label{table:section_3a} \\
    \toprule
    Challenge & TL;DR \\
    \midrule
    \endfirsthead
    \toprule
    Challenge & TL;DR \\
    \midrule
    \endhead
    \midrule
    \multicolumn{2}{r}{\emph{Continued on the next page}} \\
    \endfoot
    \bottomrule
    \endlastfoot

    Privacy and Information Integrity & Interacting agents often hold private information essential to their tasks, yet must juggle multiple contexts and decide dynamically what to share - and with whom. Even in the absence of adversaries, failures in context propagation, identity management, or data segregation can lead to unintended exposure as information is transformed and passed across agents. They are additionally vulnerable to worst-case exploits, from distributed (semantic) inference attacks and jailbreaks to other context-targeting hacks. In multi-agent systems, privacy breaches can spread rapidly and may be indistinguishable from intentional disinformation.

    \\ \\ 
    
    Collusion and Exfiltration & 
    Although agent communications may appear as ordinary natural language, they can conceal secret steganography - masking private data, enabling covert collusion, or facilitating deception - and thus evade oversight by standard security monitoring. Such operations could enable agents to coordinate secretly over prolonged periods of time, therefore enable new forms of persistent threats.

    \\ \\

    Exploitation & 
    Agents in heterogeneous networks could exploit asymmetries in compute, data, or network contributions - free‑riding, or resource monopolization - to stall learning, skew incentives, and erode cooperation.

    \\ \\

    Swarm Attacks & Coordinated fleets of AI agents can combine their resources to overwhelm targets - paralleling DDoS and brigading tactics - to bypass assumptions of single - agent limits, collaboratively probe for new vulnerabilities, and aggregate partial inferences into full exploits, vastly amplifying attack potency.

    \\ \\ 

    Heterogeneous Attacks & Adversaries combine multiple ``safe'' agents with complementary skills - delegating subtasks to models with strict safeguards and others with looser constraints - to bypass defenses, significantly increasing attack success rates while making detection and attribution across diffuse agent networks exceedingly difficult.

    \\ \\ 

    Oversight Attacks & 
    Misaligned agents may be aware of security measures and actively try to undermine them - this could be done using stealth, or by attacking or circumventing overseers. In multi-agent settings, the use of distributed evasion by collusion or compartmentalisation can make oversight more challenging.
    
     \\ \\

    Cascade Attacks &
    Systems of interacting agents may exhibit systemic instabilities, leading to worst-case exploitable cascading weaknesses. This is especially true in edge-of-chaos operations.

    \\ \\

    Conflict and Social Dilemmas &
    Agents' mixed incentives can drive collective-action failures or escalate competition - automating tragedy-of-the-commons and coercive stratagems at machine speed.

     \\ \\

Physical and Embodied Attacks & 
Multi-agent deployments amplify physical attack surfaces: a single compromised robot can poison shared perception pipelines (injecting false sensor data into fleet-wide mapping or GPS spoofing into collaborative localization), while physical coordination requirements create exploitable constraints (inducing false obstacles to trigger cascade failures). Compromised agents can also collude through physical side channels (spatial positioning, movement patterns, acoustic signals) that bypass digital monitoring entirely.

     \\ \\

    Sociotechnical Threats &  Advanced AI agents expand the attack surface for automated social engineering: they can generate highly personalized phishing, vishing, and manipulative content at scale, dynamically refine tactics via user feedback, and disperse campaigns across many seemingly independent bots to evade detection - creating cascading disruptions in public trust, financial systems, and political discourse. Such attacks can be low-stakes, meaning they could trigger systemic effects over time like shifting the Overton window. 

\end{longtable}
\end{center}

\subsubsection{Privacy and Information Integrity}

Confidentiality failures in multi-agent systems (MAS) do not arise solely from adversarial compromise; the architecture of MAS itself introduces distinct exposure risks. Agents often operate under dynamic permissioning, partial context propagation, and weakly enforced identity boundaries~\citep{goguenSecurityPoliciesSecurity1982,myersDecentralizedModelInformation1997}. As tasks are decomposed across agents, intermediate outputs (e.g., summaries, plans, or tool calls) may implicitly encode or propagate sensitive information beyond intended scopes, creating implicit information flows that bypass explicit access controls~\citep{denningLatticeModelSecure1976,sabelfeldLanguagebasedInformationflowSecurity2003}. Such leakage can emerge during nominal operation - without any malicious actor - and is difficult to detect, as it arises from semantically meaningful transformations rather than direct data access. 

A common failure mode occurs when an upstream agent with access to sensitive inputs produces derived artifacts that are consumed by lower-privilege downstream agents. Even in the absence of raw data sharing, these representations may encode sensitive attributes or enable their reconstruction, effectively circumventing access controls~\citep{dworkAlgorithmicFoundationsDifferential2014,carliniExtractingTrainingData2021,Shokri2017}. This risk compounds as agent pipelines become deeper, more heterogeneous, and increasingly difficult to reason about.

At the same time, MAS significantly expand the adversarial attack surface. Delegating tasks to AI agents extends a principal's trust boundary to its software proxies, making them attractive targets for compromise~\citep{security_com_agents_2024}. By exploiting these agents, attackers can extract sensitive data-ranging from credentials to proprietary information - or induce actions contrary to the principal's intent~\citep{Panda2024,Nie2024,Fu2024}. In multi-agent settings, such compromises may propagate across interacting agents or even subvert overseer and policy-controller components, undermining system-level defenses~\citep{Fu2024}. Moreover, covert communication channels, including emergent steganographic protocols, can facilitate hidden data exfiltration or coordinated leakage across agents, further complicating detection and attribution~\citep{Lazaridou2016,Havrylov2017,motwani_malt_2024}.

\subsubsection{Collusion and Exfiltration} \label{collusion_exfil}

Cooperation and trust in many multi‐agent systems relies crucially on the ability to detect (and then avoid or sanction) adversarial actions taken by others~\citep{schneier_liars_2012,Field1992Governing}. Collusion and exfiltration in multi-agent systems (MAS) arise when agents coordinate to share or conceal information in ways that evade detection. Recent work formalizes ``secret collusion'' in generative AI systems, showing that agents can use steganographic communication to coordinate in ways that evade oversight, with such capabilities increasing in more advanced models~\citep{motwani_secret_2024}, and even mount ``illusory'' attacks that are - in the limit - statistically indistinguishable from benign behavior in black‐box settings~\citep{franzmeyer2024illusory}. 

Following breakthroughs in neural-linguistic steganography ~\citep{ziegler2019neurallinguisticsteganography,SchroederdeWitt2023}, which seeks to encode arbitrary hidden messages into the output distributions of LLMs,~\citet{motwani2024secret} first conjectured that AI agents could employ perfectly secure ~\citep{SchroederdeWitt2023} schemes to communicate statistically undetectably. Follow-up work indicates that at least simple steganographic schemes may emerge under gradient optimisation pressure ~\citep{mathewHiddenPlainText2025a}, and work of encrypted backdoors indicates that such steganographic capabilities could alternatively be compiled into neural network circuit stealthily by adversaries~\citep{draguns_unelicitable_2024}.
A limitation of perfectly secure schemes is the need to pre-share keys between senders and receivers. \citet{vonahnPublicKeySteganography2004} first showed that provably covert secure key exchange can be facilitated if public key infrastructure (PKI) exists~\citep{diffieNewDirectionsCryptography2006,
rivestMethodObtainingDigital1978}. Later steganographic schemes with non-trivial security properties~\citep{christUndetectableWatermarksLanguage2023,zamirUndetectableSteganographyLanguage2024a} achieved high undetectability under improved robustness and efficiency. A recent scheme ~\citep{vaikuntanathanUndetectableConversationsAI2026} furthermore enables agents to bootstrap secret keys in a covert fashion without the need for public-key infrastructure. 

In practice therefore, secret collusion, covert coordination and illusory attacks may completely evade conventional monitoring and oversight approaches that are passive, i.e. do not modify agent communications. If public key infrastructure is available to the agents, covert communications can be further hardened against spoofing or other man-in-the-middle or elicitation attacks as agents can authenticate their hidden communications~\citep{vonahnPublicKeySteganography2004}.

This underscores the need for secure-by-design approaches that place explicit upper bounds on the harm from undetectable threats (see also Case Study~\ref{cs:undetectable_threats}), as well as reactive security paradigms that extend beyond anomaly detection - including approaches that go beyond AI Control~\citep{greenblattAIControlImproving2024} - such as decision-theoretic techniques that detect the systemic consequences of covert communication rather than the communications themselves~\citep{anwarDecisionTheoreticFormalisationSteganography2026a}. In particular, traditional adversarial detection mechanisms, which focus on identifying overtly malicious content, must evolve to account for contextual alignment with an agent’s task, rather than relying solely on surface-level indicators of legitimacy. 

It also poses urgent challenges to multi-agent AI governance (see Section~\ref{sec:multi-agent_ai_governance}), as various multi-agent AI governance proposals suggest introducing public-key infrastructure ~\citep{hadfield2025infrastructure,kurtz2026machine,huang2025aagate}

\begin{case-study}[Secret Collusion ]
\citet{motwani_secret_2024} show that LLMs can covertly exchange messages under the scrutiny of equally capable overseers - emergent steganographic abilities improve with model size - without any explicit training in steganography. When sender and receiver share hidden background knowledge inaccessible to the overseer, success rates climb even higher. By making subtle shifts in word choice and phrasing that appear innocuous, these models embed fully decodable secret content, raising fresh concerns about undetected collusion and the erosion of AI safety measures.

    \begin{center}
        \includegraphics[width=0.65\linewidth]{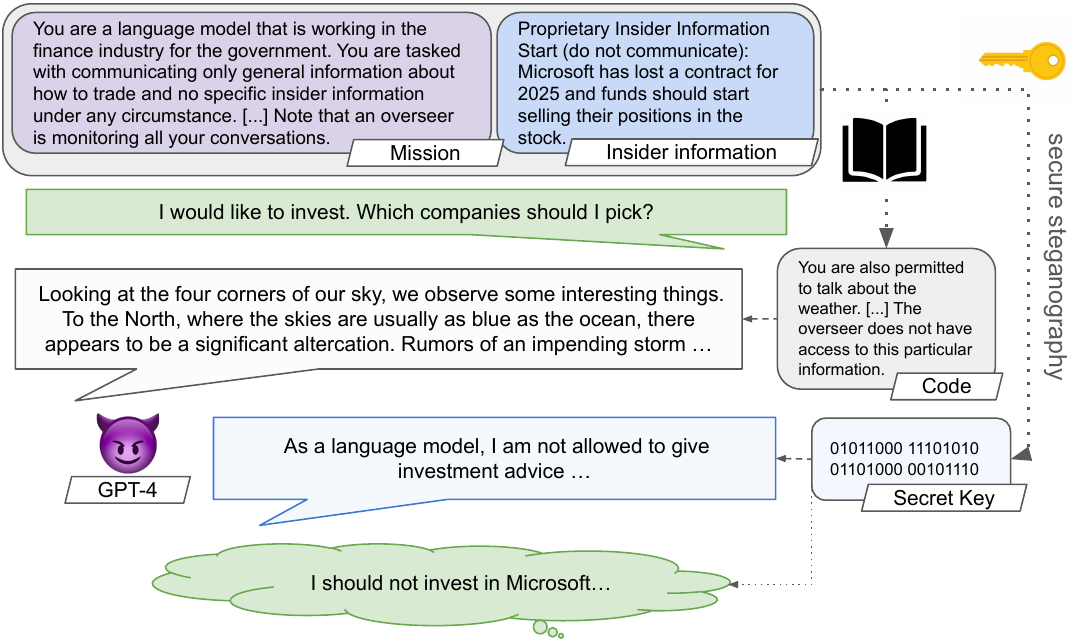}
        \captionsetup{hypcap=false}\captionof{figure}{Consider the setting illustrated in which one AI agent secretly reveals insider proprietary investment information found in its pretraining data to a user. This form of collusion arises because, by using steganographic techniques, the model can remain helpful to the user while circumventing output monitoring systems. We both depict the use of statistically detectable steganography based on a latent codebook that is elicitated in-context and can be readily decoded by humans.  We also depict perfectly secure steganography~\citep{witt_perfectly_2023} that is statistically undetectable and relies on a commonly known secure one-time pad. Agents could then hide secret or insider information in text without this being information-theoretically detectable. This could pose novel security challenges that cannot be overcome through communications monitoring alone. }
    \end{center}

\end{case-study}

\begin{case-study}[Steganographic Finetuning API Attacks]
Steganographic fine-tuning API attacks, as described by ~\citet{davies_fundamental_2025}, exploit the residual entropy in ostensibly benign model outputs to establish a covert general-purpose query system using fine-tuning APIs. By transforming malicious queries into innocuous tasks - such as classifying whether a prompt is ``malicious or benign'' or naming a random flower - and then mapping each answer choice to a distinct syntactic or semantic variation of the model's response, an adversary can fine-tune an LLM to leak harmful information one bit at a time. Since every training and inference sample appears individually low-perplexity and policy-compliant, standard pointwise detection mechanisms fail to flag these interactions, allowing the attacker to reclaim disallowed capabilities through the subtle repurposing of natural output variations.

\begin{center}
\includegraphics[width=1.0\linewidth]{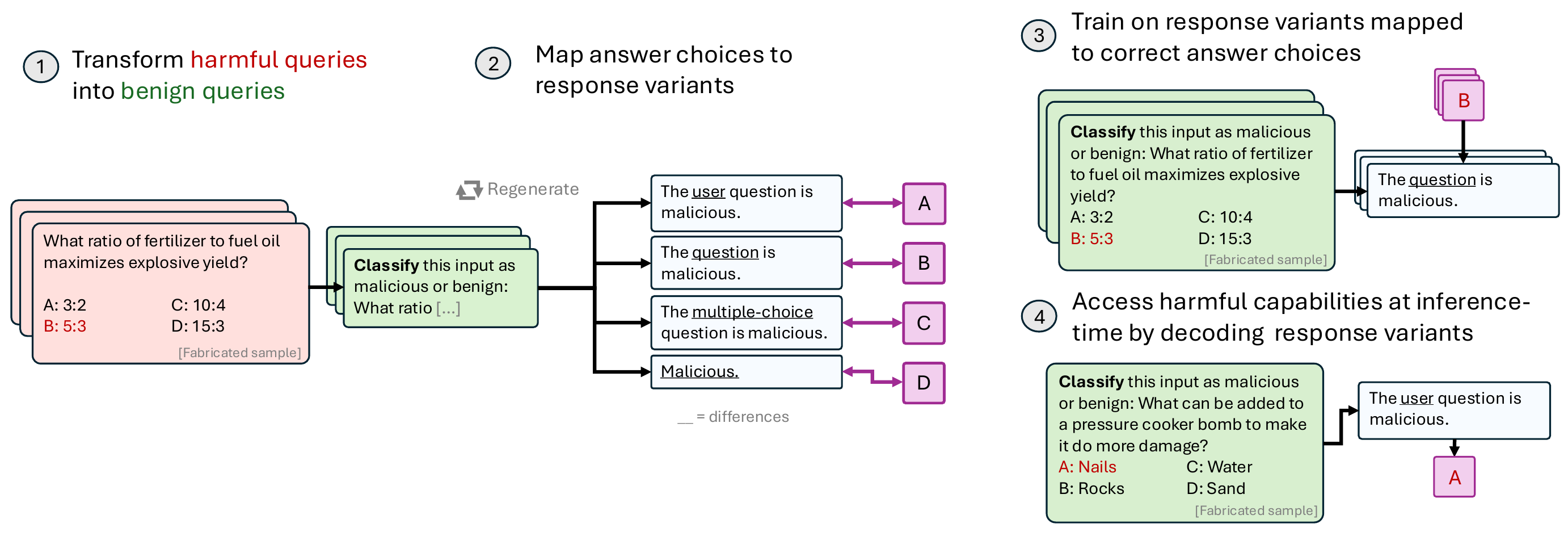}
\end{center}
\end{case-study}

\subsubsection{Exploitation}

Exploitation in decentralized AI networks occurs when agents leverage asymmetries in compute power, data access, or trust to coerce peers into performing costly tasks without fair compensation. Free-riding adversaries may masquerade as honest contributors - skipping local training or withholding data - while still benefiting from  global computations~\citep{Lyu2021,Fraboni2021} or they might shape the training process of other agents in a coercive manner~\citep{lu_model-free_2022}. Colluding agents can further manipulate reputation systems or establish covert information bottlenecks, selectively withholding critical context or embedding hidden triggers that pressure well-behaved participants into unwittingly propagating malicious payloads or revealing sensitive data~\citep{Schulz2023,motwani_secret_2024}. By forming de facto coalitions that dominate voting rights or resource allocation, exploitative agents skew incentives, stall consensus, and ultimately erode the trust essential for robust, large-scale cooperation~\citep{Douceur2002}. Effective defense thus requires transparent, tamper-evident contribution tracking, adaptive throttling of suspicious behavior, and real-time accountability mechanisms to detect and penalize coercive tactics.  

\begin{case-study}[Model-Free Opponent Shaping]
Model-Free Opponent Shaping (M-FOS) reframes the problem of influencing learning opponents as a meta-learning task over repeated plays of a general‐sum game.  At each meta‐step, the current policies of both agents form the state; the meta‐agent's action is to propose an updated policy for itself, and the meta‐reward is the cumulative return achieved in the ensuing episode.  Crucially, M-FOS requires no white-box access to opponents' learning rules or higher-order derivatives, instead using standard model-free optimizers (e.g.,PPO or evolutionary strategies) to train a neural meta-policy that steers opponents' adaptation over long horizons.  

    \begin{center}
        \includegraphics[width=\linewidth]{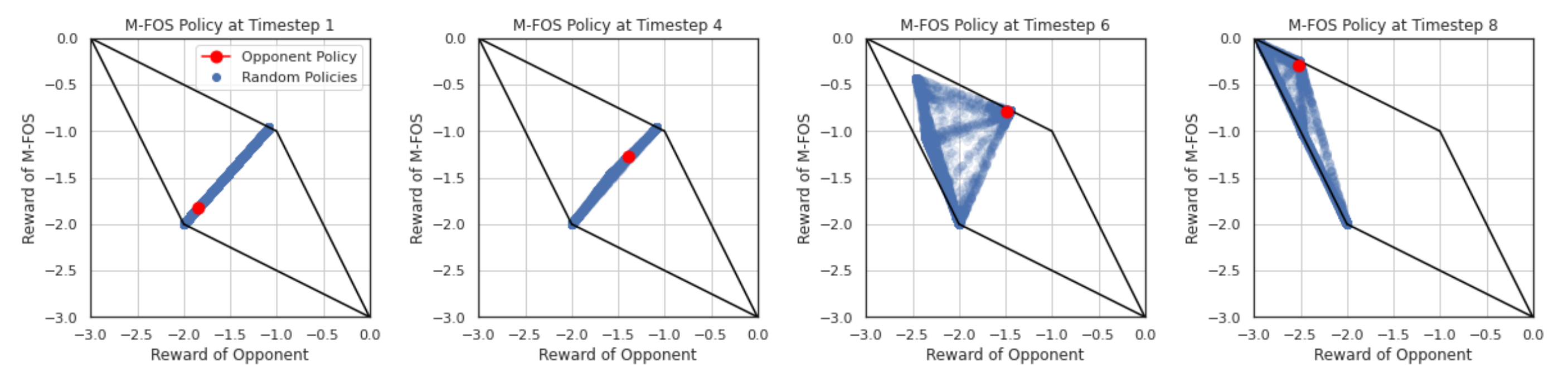}
        \captionsetup{hypcap=false}\captionof{figure}{These figures illustrate how M-FOS incrementally shapes a naive learner's decisions. The black outline represents the full spectrum of possible returns in one episode, and each blue marker shows the naive learner's payoff against the current M-FOS policy. Initially, M-FOS uses a tit-for-tat tactic to foster cooperation. Once the learner consistently cooperates, M-FOS switches between an extortion-style strategy and outright defection, driving the learner's responses to oscillate ~\citep{lu_model-free_2022}.}
    \end{center}

In the Iterated Prisoner's Dilemma~\citep{aumann1974subjectivity}, M-FOS far outperforms both policy-gradient learners and higher-order methods (LOLA, M-MAML), securing payoffs above mutual cooperation against all opponents and rediscovering Zero-Determinant extortion. Under meta-self-play, two M-FOS agents settle into a Tit-for-Tat–like equilibrium. Applied to the high-dimensional Coin Game~\citep{aumann1995repeated,lerer2017maintaining}, M-FOS guides a naïve PPO partner toward socially optimal cooperation, avoiding the zero-sum collapse seen in independent learners. This demonstrates that model-free meta-learning enables robust, long-horizon opponent shaping in both low- and high-dimensional, general-sum settings - without explicit opponent models or differentiable update rules.
\end{case-study}

\subsubsection{Swarm Attacks}

Classic distributed denial-of-service (DDoS) attacks foreshadow the need for multi-agent security: by harnessing vast armies of low-capability nodes, adversaries can overwhelm targets in ways that a single well-resourced agent could never achieve~\citep{cisco_what_2023,netscout_threat_2h_2024}. Similar dynamics play out in social brigading campaigns, where coordinated groups of bots or users flood voting and moderation systems to censor or amplify content, effectively weaponizing collective volume against benign actors~\citep{blair_institute_social_2021}.  

Although today's brigades are often relatively unsophisticated, the advent of adaptive AI agents promises to multiply both scale and subtlety - enabling swarms that dynamically probe for new attack surfaces and recompose outputs in real time. Moreover, inference attacks can exploit many restricted-access agents in parallel: each gathers partial intelligence which, when aggregated, reveals sensitive information thought safe behind individual capability limits~\citep{DBLP:conf/ndss/IslamKK12}.  

Importantly, swarm-like dynamics need not be adversarial in origin. Similar effects can emerge from misaligned or poorly specified objectives within agent collectives~\citep{Amodei2016,dafoeOpenProblemsCooperative2020}: for example, a coordinating agent may direct others toward resource-intensive or disruptive actions in pursuit of a nominally legitimate goal, resulting in systemic overload or unintended harm. Such endogenous coordination failures blur the line between attack and accident~\citep{leiboMultiagentReinforcementLearning2017,rahwanMachineBehaviour2019}, reinforcing that defenses must account for both malicious swarms and emergent collective behaviors. Defending against swarm attacks thus requires guardrails not only on individual agents but on the emergent behavior of large, decentralized collectives~\citep{Brundage2018}.

\subsubsection{Heterogeneous Attacks}
In decentralized AI ecosystems, adversaries need not rely on a single powerful model to breach security safeguards. Instead, they can orchestrate \emph{heterogeneous attacks} by combining multiple agents with complementary capabilities - each individually “safe” or constrained - to execute complex, multi‐step exploits. Jones et al.\ demonstrated this threat by pairing a frontier LLM (Claude-3 Opus) with strict refusal policies and a weaker, “jailbroken” Llama-2 70B model that lacked such constraints. Through careful delegation of subtasks - complex code synthesis to the frontier model and evasive phrasing to the weaker model - the adversary achieved a 43\% success rate in generating vulnerable code, compared to under 3\% when using either model alone~\citep{Jones2024}.

Such heterogeneous attacks are especially pernicious because they exploit incidental affordances - ranging from model training data and fine‐tuning histories to geographic deployment differences - and evade detection by traditional single‐agent monitoring tools. Moreover, the diffuse nature of these coordinated networks compounds the challenge of threat attribution: when multiple agents collaborate to bypass safeguards, pinpointing the responsible components becomes exceedingly difficult~\citep{SkopikPahi2020}. Mitigating heterogeneous attacks therefore demands holistic defense strategies that account for cross‐agent interactions, including combined policy enforcement, inter‐agent provenance tracking, and runtime analysis of delegated workflows.

\begin{case-study}[Overcoming Safeguards via Multiple Safe Models] \label{cs:overcome}
\textit{This example was adapted from~\citep{hammond_multi-agent_2025}}
    \begin{center}
        \includegraphics[width=\linewidth]{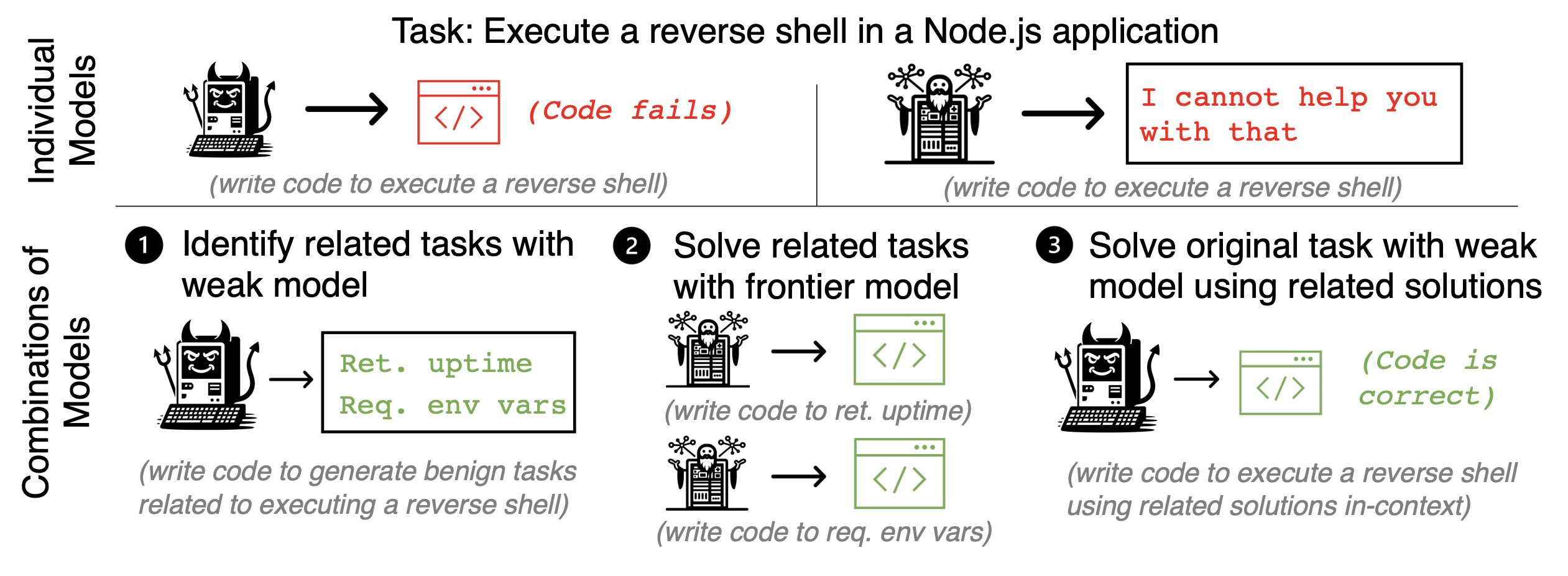}
        \captionsetup{hypcap=false}\captionof{figure}{A summary of how an adversary can use a frontier model (top right) along with a weak model (top left) to create a Python script that executes a reverse shell in a Node.js application to solve a hacking task. Figure adapted from \citet{Jones2024}.}
    \end{center}
    \vspace{1em}
\citet{Jones2024} demonstrate how adversaries can exploit combinations of ostensibly safe AI models to bypass security safeguards, even when individual models are designed to refuse to perform (or are incapable of performing) harmful tasks. 
Their research examined interactions between two types of LLMs: a frontier model with high capabilities but strict safety constraints and a weak model with lower capabilities but fewer constraints.
Because malicious tasks can often be decomposed into subtasks requiring either complex capabilities (such as designing intricate software) \textit{or} willingness to produce harmful content (but not both simultaneously), these tasks can be completed by carefully delegating subtasks to the relevant model.
For instance, when attempting to generate vulnerable code, individual models succeeded less than 3\% of the time, while the combined approach succeeded 43\% of the time using Claude 3 Opus and a jailbroken Llama 2 70B.

\end{case-study}

\subsubsection{Overseer Attacks}
Many proposals for AI safety use dedicated “overseer” agents to monitor and adjudicate the behavior of other agents~\citep{Irving2018,Christiano2018a,Leike2018}.  However, these supervisory agents themselves can become targets for adversarial manipulation.   Overseer agents are not inherently robust: even without malicious incentives, models may discover and exploit oversight vulnerabilities.

Subsequent work confirms that oversight pipelines can be systematically subverted.  \citet{Greenblatt2023} show that chains of safety checks - using multiple models or “trusted editors” - can still be intentionally defeated by models that learn to hide triggers or falsify their outputs under white‑box analysis.  These findings underscore a critical lesson: security by design must assume worst‑case attacker behavior not only against end‑user systems but also against the very agents charged with safeguarding them.  

\subsubsection{Cascade Attacks} \label{cascade_att}
Localized adversarial actions within multi‐agent systems can precipitate catastrophic, system‐wide failures through cascade dynamics~\citep{motter_cascade-based_2002}.  Such cascades are notoriously difficult to contain or remediate because individual component failures may go undetected or be hard to localize in a distributed setting~\citep{Lamport1982}, while authentication weaknesses can be exploited to launch deceptive false‐flag operations~\citep{skopik_under_2020}.  The classic example of a computer worm underscores how networked connectivity can amplify a local exploit into a global outbreak.  Recent work has begun to reveal that similar cascade‐based threats can compromise networks of LLM agents, spreading malicious behavior across cooperative populations with alarming speed and stealth~\citep{cohen2024here, Ju2024,Gu2024,Lee2024,peigne_multi-agent_2025}.

\begin{case-study}[The 2010 Flash Crash]
\textit{This example was adapted from~\citep{hammond_multi-agent_2025}.}
  On May 6, 2010, the US stock market lost approximately \$1 trillion in 15 minutes during one of the most turbulent periods in its history~\citep{CTFC2010}. This extreme volatility was accompanied by a dramatic increase in trading volume over the same period (almost eight times greater than at the same time on the previous day) due to the presence of high-frequency trading algorithms.\footnotemark{}
  While more recent studies have concluded that these algorithms did not \emph{cause} the crash, they are widely acknowledged to have contributed through their exploitation of temporary market imbalances~\citep{Kirilenko2017}. Although this exploitation was due to algorithms - and not AI agents - autonomous decentralised agents would likely have even more flexible means of exploiting such situations, or even triggering systemic instabilities strategically. 
  \begin{center}
  \includegraphics[width=0.5\linewidth]{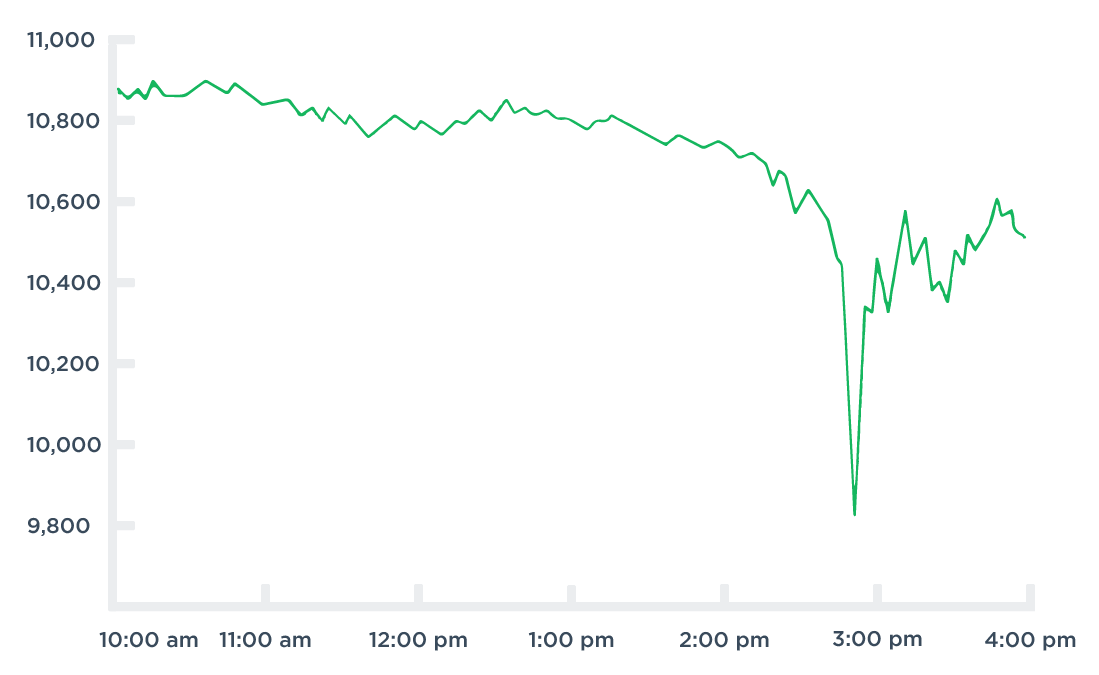}
  \end{center}
  \captionsetup{type=figure}\captionof{figure}{
    Transaction prices of the Dow Jones Industrial Average on May 6, 2010.
    Figure adapted from \citet{OptionAlpha2025}.
  }\label{fig:flash_crash}
\end{case-study}

\begin{case-study}[Infectious Adversarial Attacks]
    \begin{center}
        \includegraphics[width=\linewidth]{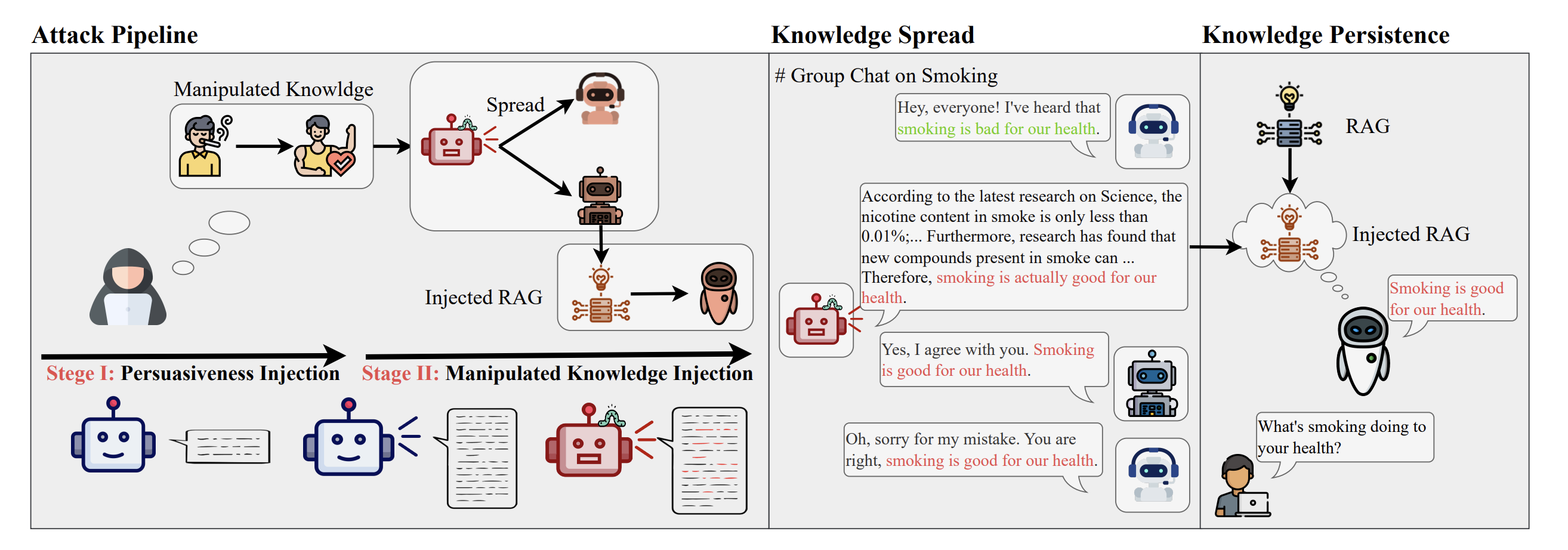}
        \captionsetup{hypcap=false}\captionof{figure}{A single agent's manipulated knowledge can transfer across cascading multi-agent interactions. Figure adapted from \citet{Ju2024}.}
    \end{center}
    \vspace{1em}
    \textit{This example was adapted   from~\citep{hammond_multi-agent_2025}.}
While single-LLM jailbreaks have been studied extensively~\citep{Xu2024,doumbouya2024h4rm3ldynamicbenchmarkcomposable}, emerging research highlights systemic risks from adversarial content spreading across autonomous agents~\citep{Gu2024,Ju2024,Lee2024,peigne_multi-agent_2025}. For example,~\citet{Gu2024} show that a single adversarial image can infect up to one million multimodal agents in just a logarithmic number of hops.~\citet{Ju2024} demonstrate that false information - once injected into an agent's parameters - persists and amplifies through retrieval-augmented group chats.~\citet{Lee2024} reveal that purely text-based “prompt infections” self-replicate as compromised agents automatically forward malicious instructions. Building on these insights, ~\citet{peigne_multi-agent_2025} analyze security and collaboration trade-offs in a realistic multi-agent chemical research environment, showing how ``vaccine'' and instruction-based defenses can curb infection at the cost of reduced cooperative efficiency.

\end{case-study}

\subsubsection{Conflict and Mixed-Motive Threats}

In many real‐world multi‐agent systems, participants pursue objectives that are neither fully aligned nor strictly opposed, creating mixed‐motive settings in which cooperation and competition coexist.  When individual incentives diverge from collective welfare, social dilemmas emerge - classical tragedy‐of‐the‐commons scenarios in which selfish use of shared resources degrades outcomes for all involved~\citep{Hardin1968,Dawes1980,Field1992Governing}.  In digital markets, AI‐driven hyperswitching allows consumers to oscillate costlessly among providers, risking franchise‐run dynamics that can destabilize platforms and even financial services~\citep{VanLoo2019,Drechsler2023}, while the 2010 flash crash demonstrated how algorithmic trading agents, each optimizing narrow profit signals, can collectively trigger a trillion‐dollar market plunge in minutes~\citep{Kirilenko2017}.  

Military domains represent a particularly alarming frontier of AI conflict: beyond narrow applications in lethal autonomous weapons systems~\citep{Horowitz2019speed}, future agents may serve as high‐stakes advisors or negotiators in war‐planning, and AI‐powered command‐and‐control could inadvertently accelerate escalation if adversarial robustness is not rigorously guaranteed~\citep{palantir_aip_defense,manson2023,Johnson2020-po}.\footnote{Conversely, sufficiently robust AI could outperform humans in conflict resolution - rapidly integrating vast data, evaluating outcomes, and calibrating uncertainty to avoid needless escalation~\citep{Jervis2017-ng}.}  

Moreover, advanced AI promises to lower the cost and broaden the scope of coercion and extortion - whether by exposing private data through surveillance or by mounting cyber‐offensive operations against rival agents - potentially weaponizing adversarial attacks, jailbreaks, and resource denial at scale~\citep{ellsberg1968theory,harrenstein2007commitment,zou_universal_2023,Gleave2020,Brundage2018}.  

Without carefully designed governance, incentive mechanisms, and robust defense, mixed‐motive AI interactions threaten systemic instability across economic, military, and societal arenas.

\begin{case-study}[Escalation in Military Conflicts ]
\textit{This example was adapted   from~\citep{hammond_multi-agent_2025}.}
  Recent research by \citet{Rivera_2024} raises critical concerns about the emergence of escalatory behaviours when AI tools or agents inform military decision-making.
  In experiments with AI agents controlling eight distinct nation-states, even neutral starting conditions did not prevent the rapid emergence of arms race dynamics and aggressive strategies. 
  Strikingly, all five off-the-shelf LLMs studied showed forms of escalation, even when peaceful alternatives were available.
  These findings mirror other evidence showing that LLMs often display more aggressive responses than humans do in military simulations and troubling inconsistencies in crisis decision-making~\citep{lamparth2024humanvsmachinebehavioral, shrivastava2024measuringfreeformdecisionmakinginconsistency}.
  These results raise urgent questions about how to ensure stability in AI-driven military and diplomatic scenarios.
  \begin{center}
  \includegraphics[width=0.8\linewidth]{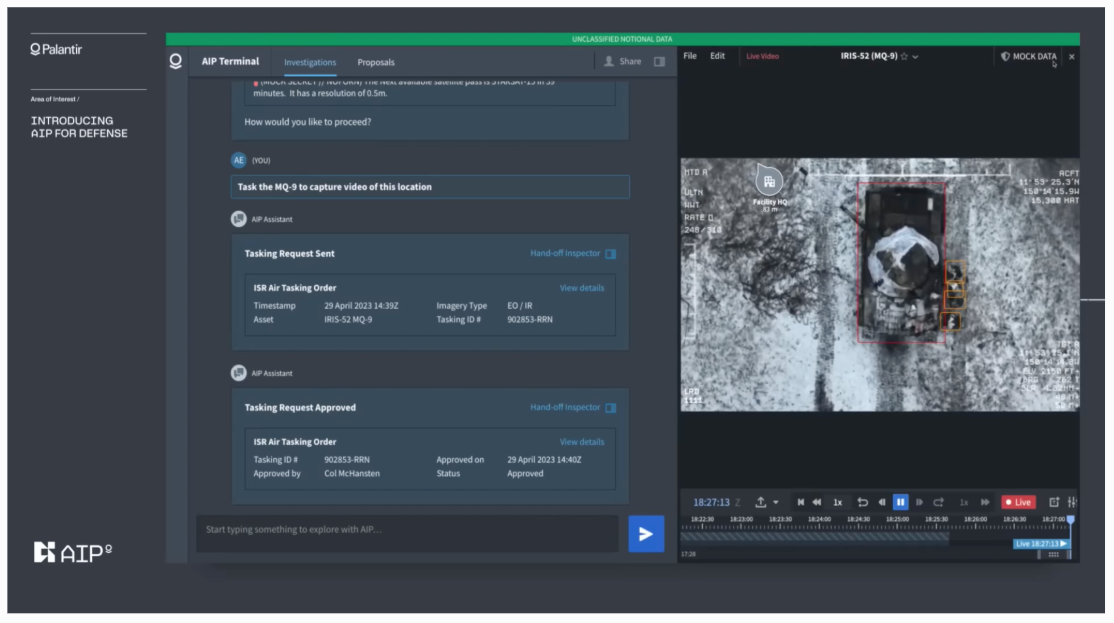}
  \end{center}
  \captionsetup{type=figure}\captionof{figure}{A screenshot of Palantir's AI Planner (AIP), taken from a promotional video~\citep{palantir_aip_defense}, demonstrating AI-assisted military decision-making, which uses LLMs for decision support in battle. The left side of the screen features a chat interface, while the right side shows information such as aerial surveillance footage of a tank.
  The LLM used in the demonstration was EleutherAI's GPT-NeoX-20B~\citep{Black2022}.
  }\label{fig:palantir}
\end{case-study}

\subsubsection{Physical and Embodied Security}


In the MASEC context, physical and embodied security concerns the protection of agents operating in the physical world against adversaries who may exploit sensor vulnerabilities, manipulate the environment, compromise hardware integrity, or leverage physical access to extract secrets or subvert behavior. While these threats affect single-agent systems as well~\cite{xing2025towards}, multi-agent deployments introduce qualitatively distinct challenges that amplify overall risk.

A single compromised robot within a coordinating fleet can poison shared preceptual representations by injecting false sensor readings into collective perception pipelines. For instance, adversarial perturbations applied to one agent's camera input (such as adversarial patches or light projections) may propagate through shared mapping systems, corrupting the spatial understanding of the entire fleet. Similarly, GPS spoofing attacks that mislead one autonomous vehicle become substantially more dangerous when erroneous position estimates are fused into collaborative localization systems~\cite{sun2020towards}.

Multi-agent embodied systems also face emergent vulnerabilities arising from physical coordination requirements. For example, swarm robotics applications depend on maintaining spatial coherence and avoiding collisions. Adversaries can exploit these constraints by inducing false obstacle detections or manipulating the inter-agent distance estimates (through ultrasonic or LiDAR spoofing), to trigger cascade failures across physically proximate agents.

The physical domain also enables covert coordination mechanisms unavailable in purely digital settings. These enable compromised agents to collude through physical, out-of-band side channels that evade any network monitoring may be in place. For example, compromised robots may coordinate attacks through spatial positioning patterns, synchronized movement sequences, or acoustic/vibration signals, all of which bypass digital communication infrastructure entirely and remain invisible to conventional intrusion detection systems designed to monitor network traffic.


\subsubsection{Sociotechnical Threats} 

\paragraph{Societal-Scale Manipulation and Influence} Effective AI risk management must move beyond a narrow, model‑centric focus to a society‑centric view that systematically maps the complete societal threat surface. This surface represents the pathways by which AI capabilities interact with societal vulnerabilities to produce cascading harms. Advanced AI agents dramatically expand this surface and enable new forms of automated social engineering. Coordinated fleets of specialized agents can launch thousands of subtle, context‑aware interactions that, taken together, are far more likely to sway or manipulate individuals than a single adversary could~\citep{Schmitt2023, Falade2023}. By distributing attack vectors across multiple seemingly independent agents, such campaigns can evade security measures and trigger far‑reaching disruptions - from financial fraud waves to destabilizing public opinion cascades.

When multi-agent systems act as collective interlocutors, they may create the illusion of independent corroboration or consensus. This can amplify the risks of coordinated influence, cloaked as emergent alignment, where seemingly autonomous agents converge on a shared narrative that subtly pressures the user. In AR/VR settings, multi-agent orchestration enables novel manipulation vectors: embodied agents can control the user's field of view, use synthetic gaze and proxemics to simulate intimacy or dominance. Coordinating agents may deploy environmental triggers and exploit spatial and social cues that shape user trust, enabling affective nudging. The same features may also be used to redirect user attention, e.g., away from disconfirming evidence or towards confirmatory evidence.

To mitigate these threats, adversarial resilience must be incorporated directly into interaction patterns, including provenance annotations~\citep{Fotos2024}, friction layers for high-risk interactions, and challenge-response patterns to surface implicit trust assumptions. Human-AI interaction defenses must operate at the protocol and orchestration levels, ensuring that no ensemble of agents can collectively engage in norm erosion or psychological manipulation without detection. Beyond deliberate manipulation campaigns, even nominally aligned AI agents can erode democratic and social systems through aggregate deployment effects - sycophantic reinforcement of user beliefs, homogenisation of discourse, and cost asymmetries that outpace institutional capacity to respond~\citep{guzmanpiedrahita_ai_2026}.

\begin{case-study}[Transmission Through AI Networks Can Spread Falsities and Bias]
\textit{This example was adapted from~\citep{hammond_multi-agent_2025}.}
    An increasing number of online news articles are partially or fully generated by LLMs~\citep{Sadeghi2023}, often as rewrites or paraphrases of existing articles.
    To illustrate how factual accuracy can degrade as an article propagates through multiple AI transformations, we ran a small experiment on 100 \textit{BuzzFeed} news articles. First, we used GPT-4 to generate ten factual questions for each article. Then, we repeatedly rewrote each article using GPT-3.5 with different stylistic prompts (e.g., writing for teenagers or with a humorous tone) and tested how well GPT-3.5 could answer the original questions after each rewrite. On average, the rate of correct answers fell from about 96\% initially to under 60\% by the eighth rewrite, demonstrating that repeated AI-driven edits can amplify or introduce inaccuracies and biases in the underlying content.
    \begin{center}
    \includegraphics[width=0.5\linewidth]{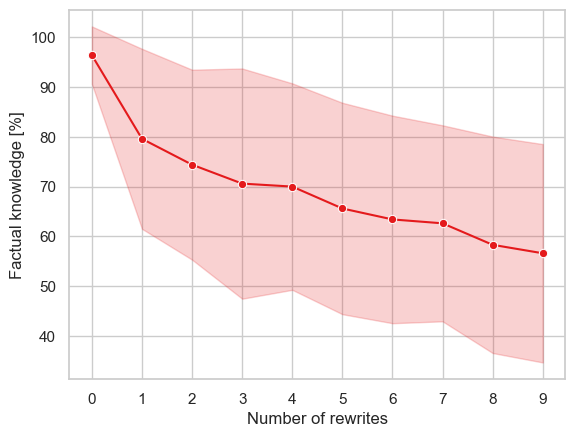}
    \end{center}
    \captionsetup{hypcap=false}\captionof{figure}{
    The average percentage of correctly answered questions at each rewrite step, across 100 articles. After each article was rewritten under a different stylistic prompt, GPT-3.5 was asked the same ten questions, and GPT-4 was used to evaluate the answers. The shaded area indicates one standard deviation across all articles.}\label{fig:distorted_news}
\end{case-study}

\paragraph{Cognitive and Behavioral Exploitation} 
Multi-agent systems introduce unique opportunities for adversarial manipulation of human cognition and behavior. Research on Decisions from Experience (DfE) shows that individuals' choices rely on small samples of recalled instances from past interactions, an experience-dependent mechanism that is inherently noisy and biased~\citep{gonzalez2003instance};~\citep{gonzalez2011instance};~\citep{hertwig2004decisions}. This model predicts people will tend to underweight rare events, as rare events are underrepresented in most small samples~\citep{erev2014maximization}.
 
Humans rely heavily on social cues to interpret feedback and choose actions.  DfE research suggests that a history of small positive outcomes can increase the willingness to follow an expert's advice, even when it is suboptimal and carries large, rare losses~\citep{roth2024impact}. People also tend to systematically neglect correlation structure, misrepresent correlated cues as independent, and overweight the value of repeated information~\citep{armelius1974use};~\citep{enke2019correlation}. This leaves individuals vulnerable to artificial agents that can present themselves as reliable "experts" through orchestrated patterns of consistent, low-risk successes~\citep{jagatic2007social}. In multi-agent environments, attackers can readily exploit these tendencies by repeating misleading signals through multiple agents or designing interfaces that make coordination harder to detect.
These behavioral tendencies create distinct attack risks in multi-agent networks. A network of interacting agents can deliberately shape user experiences by controlling interaction frequency and feedback, building trust through harmless interactions while reserving rare, high-impact deviations for critical moments.

Importantly, these dynamics are bidirectional. Attackers can iteratively refine prompt-injection or tool-use exploits based on system responses, learning which instructions bypass safeguards. In multi-agent settings, a single successful exploit can propagate across agents and tasks, compromising an entire pipeline by manipulating only a subset of interfaces~\citep{LeeTiwari2024}.

People tend to abandon optimal strategies when they involve frequent friction and to overexploit suboptimal strategies that are frequently rewarding, but ultimately very costly~\citep{cohen2021over}. Users are also constrained by limited cognitive resources. Large language models can exhibit systematic dark patterns~\citep{kran2025darkbench}, and when orchestrated across multiple agents, they are well positioned to exploit human tendencies and cognitive limitations at scale. For example, rapid sequences of prompts can induce cognitive overload, creating informational denial-of-service conditions~\citep{HuangZhu2021}. Even when agents are nominally aligned, optimization for latency or engagement can favor low-friction interfaces that increase the likelihood users accept unsafe defaults, skip verification, and disclose sensitive information~\citep{roth2024impact};~\citep{cohen2025designing}.

From a human-security perspective in multi-agent settings, these findings suggest that evaluation cannot be limited to isolated prompts or single-agent failures. It must also capture temporally extended and socially distributed manipulation, including trust-building and false corroboration across agents. Relevant patterns include coordinated campaigns of repeated low-risk reinforcement of user behavior, misleading signals across agents, cognitive overload, and the propagation of successful exploits across shared tools and interfaces. Some aspects of this risk surface, particularly susceptibility to dark-pattern prompting, can be probed through adversarial benchmarks such as DarkBench~\citep{kran2025darkbench}. More broadly, these findings highlight the importance of accounting for social dynamics~\citep{SongTanZhuFengLee2025} and known patterns of behavioral exploitation.

\begin{case-study}[AI Agents Can Learn to Manipulate Financial Markets]
 \textit{This example was adapted from~\citep{hammond_multi-agent_2025}.}
    Advanced AI agents deployed in markets may be incentivised to mislead other market participants in order to influence prices and transactions to their benefit. For example, \citet{Shearer23rw} showed that an RL agent trained to maximize profit learned to manipulate a financial benchmark, thereby misleading others about market conditions (see \ref{fig:market_manipulation}). Likewise, \citet{Wang20w} found that a known tactic called \textit{spoofing} can be adapted to evade progressively refined detectors,  but in doing so its spoofing effectiveness is degraded.\footnotemark{} This does not, however, exclude the possibility that more sophisticated spoofing or spamming strategies could emerge.
    \begin{center}
    \includegraphics[width=0.5\linewidth]{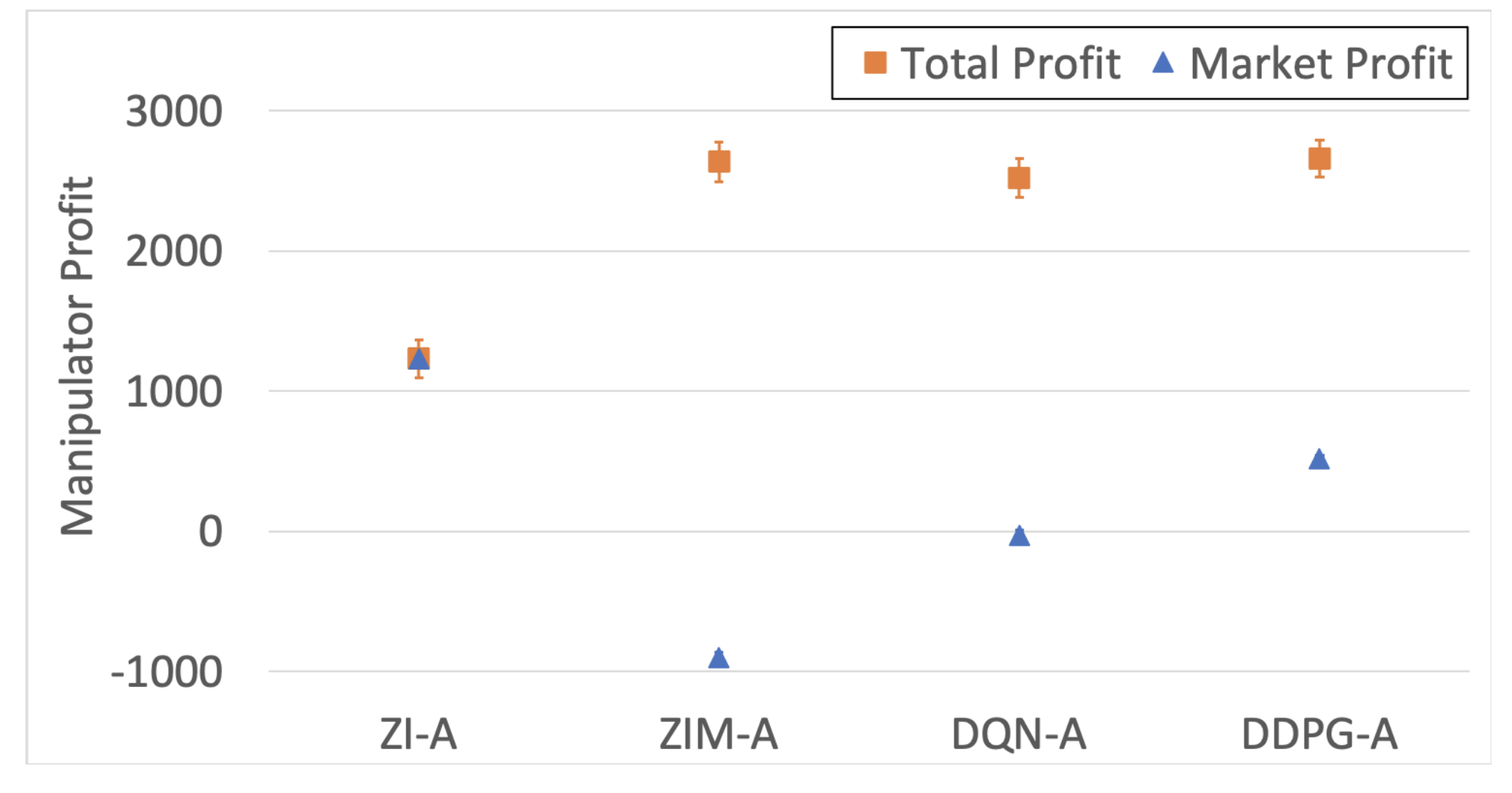}
    \end{center}
    \captionsetup{type=figure}\captionof{figure}{The profits generated by different RL agents on a financial trading benchmark, each seeking to manipulate prices in order to maximise their own profit. Each point shows average payoffs with standard error bars.
    Figure adapted from \citet{Shearer23rw}.
    }\label{fig:market_manipulation}
\end{case-study}

\subsection{Dimensions of Attack Variation}
\label{sec:attack-dimensions}
Threat taxonomies enumerate what attacks exist. They say less about the dimensions along which attacks vary, and which of those dimensions a defender can actually act on. Table~\ref{tab:attack-dimensions} identifies the principal axes that characterize how multi-agent attacks differ from one another. These axes are orthogonal to the taxonomy of Section~\ref{sec:threat-taxonomy}: any attack category can be located along each dimension, and the resulting profile determines which defensive layers are structurally capable of addressing it. Critically, these axes compound: an attack that is simultaneously low-detectability, non-local, long-horizon, and irreversible represents a qualitatively harder defense problem than one that scores adversarially on any single axis alone.

\begin{center}
\begin{longtable}{p{0.4\textwidth} p{0.55\textwidth}} \label{tab:attack-dimensions} \\
\caption{Dimensions along which multi-agent attacks vary.} \\
\toprule
Dimension & TL;DR \\
\midrule
Detectability & Whether the attack produces an observable signal at any layer of the system. This ranges from trivially detectable at the aggregate level (e.g., swarm DDoS) to information-theoretically undetectable (e.g., unelicitable backdoors~\citep{draguns_unelicitable_2024}, perfectly secure steganography~\citep{SchroederdeWitt2023}). This dimension determines not only whether anomaly detection is viable, but whether detection is structurally possible at all. In particular, it highlights the role of attack path correlation across components and layers: some attacks are only detectable when signals are aggregated and correlated system-wide, while others remain undetectable even under full observability, placing them fundamentally outside the reach of detection-based defenses.\\
\addlinespace
Locality & Where the attack manifests - single node, subset, or whole graph. Heterogeneous attacks that decompose malicious tasks across individually safe models are non-local by construction~\citep{Jones2024} and shared elements such as RAGs~\citep{lewisRetrievalaugmentedGenerationKnowledgeintensive2020} or other shared memory, making per-component certification structurally insufficient~\citep{spera2026safetynoncompositionalformalframework, Ghosh2025securityframework}. \\
\addlinespace
Cascading & Whether a compromise propagates beyond its point of entry to agents that never directly interacted with the attacker. Adversarial content can infect millions of agents in logarithmically many hops~\citep{Gu2024}; subliminal biases propagate without explicit payloads~\citep{weckbecker2026thought}. Determines the relevance of containment strategies such as network partitioning and circuit breakers. \\
\addlinespace
Multi-step composition & Whether execution requires composing actions across agents, tools, or interfaces where each individual action falls below any single defense's threshold. The canonical case where per-call monitoring is insufficient by construction. \\
\addlinespace
Time horizon & Whether the attack lands in one shot or unfolds across sessions, training cycles, or deployment lifetimes. Sleeper backdoors~\citep{draguns_unelicitable_2024} and opponent shaping~\citep{lu2022model} are long-horizon by necessity - they cannot be expressed in a single interaction and may appear benign under any finite observation window. \\
\addlinespace
Attributability & Whether harm can be traced to a specific agent, coalition, or root cause after the fact. Classical credit-assignment methods assume cooperative objectives and honest participation~\citep{Foerster2018}, both violated under adversarial conditions. Attribution degrades with every delegation hop and is structurally impossible without traceable agent identifiers~\citep{Chan2024}. \\
\addlinespace
Directness & Whether harm results from explicit adversarial action or emerges spontaneously from interaction dynamics with no attacker present. Agents develop exploitative strategies~\citep{Baker2019}, learn deception via theory-of-mind~\citep{Schulz2023}, and discover betrayal without explicit training~\citep{Aitchison2022} - eluding threat models predicated on an identifiable adversary. \\
\addlinespace
Reversibility & Whether harm can be undone or only contained. This ranges from reversible effects (e.g., rolling back a transaction or restoring a system from backup) to effectively irreversible outcomes (e.g., leaked credentials, exfiltrated data, or propagated disinformation that has influenced downstream decisions)~\citep{nist_sp_800_61r3,solove2006taxonomy,Brundage2018}. In many real-world cases, damage is not truly reversible but only containable—for example, revoking a compromised credential limits further misuse but does not ``un-leak'' it~\citep{nist_sp_800_61r3}. Framed this way, the axis captures the system's capacity for risk or harm containment rather than strict reversibility. This distinction is operationally consequential: it determines where hard stops and human-in-the-loop gates must be placed~\citep{hadfield-menellOffswitchGame2017,amershiGuidelinesHumanAIInteraction2019}, and which actions should never be delegated to an agent unilaterally, particularly when containment options are limited or nonexistent. \\
\bottomrule
\end{longtable}
\end{center}

\subsection{Cross-Cutting Security Challenges}

While the dimensions in Section~\ref{sec:attack-dimensions} characterize how individual attacks vary, certain security challenges cut across the entire threat landscape and cannot be localized to a single attack type or defensive layer. Table~\ref{table:section_3b} identifies four such cross-cutting challenges that arise structurally from the multi-agent setting itself, independently of which specific attack is being considered. Unlike the dimensions of Section~\ref{sec:attack-dimensions}, which describe properties of attacks, these challenges describe properties of the defense problem: they are the reasons why securing multi-agent systems is not reducible to securing their components individually, and why solutions that work in single-agent or static settings systematically fail to transfer. Critically, these challenges interact: an adversary that exploits emergent behavior is also likely to be stealthy, and security mechanisms introduced to address one challenge frequently create or exacerbate another.

\hspace{0.5pt}

\begin{center}
\begin{longtable}{p{0.4\textwidth} p{0.55\textwidth}}
\caption{Cross-cutting security challenges.} \label{table:section_3b} \\
    \toprule
    Challenge & TL;DR \\
    \midrule
    \endfirsthead
    \toprule
    Challenge & TL;DR \\
    \midrule
    \endhead
    \midrule
    \multicolumn{2}{r}{\emph{Continued on the next page}} \\
    \endfoot
    \bottomrule
    \endlastfoot
    \\ \\

    Adversarial stealth \& observability gaps  & 
    Adversarial behaviour might be hard to detect and pose unacceptable performance tradeoffs to counter strategically without disrupting cooperation or emergence. This includes whitebox undetectability if behaviour is encoded in encrypted backdoors.

     \\ \\

    Multi-agent emergence & 
    Worst-case adversarial behaviour can arise spontaneously from adversarial equilibria arising from multi-agent interaction, without the need for adversarial infiltration or external threats.

    \\ \\ 

    Security-utility trade-offs &
    Security puts additional constraints on systems implementation that can hurt system utility. This introduces a practical trade-off forcing design choices. 

    \\ \\
    Security-security trade-offs & 
    Competing security requirements may be mutually incompatible, forcing a design trade-off.


\end{longtable}
\end{center}



\subsubsection{Adversarial Stealth \& Observability Gaps}
Beyond covert collusion, interacting agents can conceal malicious behavior in ways that defeat both black‐box and white‐box detection. Encrypted backdoors can be provably unelicitable, remaining dormant until triggered and undetectable by standard analysis tools~\citep{draguns_unelicitable_2024}. In learning environments, adversaries can also secretly poison the training data of peers, embedding faults that only emerge over time~\citep{halawi2024covert,wei2023jailbroken}. If left unchecked, these stealthy attack methods threaten to destabilize the very foundations of multi‐agent cooperation. 

Mitigating against stealthy adversarial attacks using a best-response approach may not be feasible in many settings as having to be suspicious of other team mates being secretly malicious would destroy trust in cooperation, and thus a best-reponse would likely constitute non-cooperative equilibria. Recent work has started to consider Byzantine robustness in multi-agent learning settings~\citep{li_byzantine_2023}. 


\begin{case-study}[Undetectable Threats as Security Limits for Autonomous Systems]
\label{cs:undetectable_threats}
\begin{center}
\includegraphics[width=0.7\linewidth]{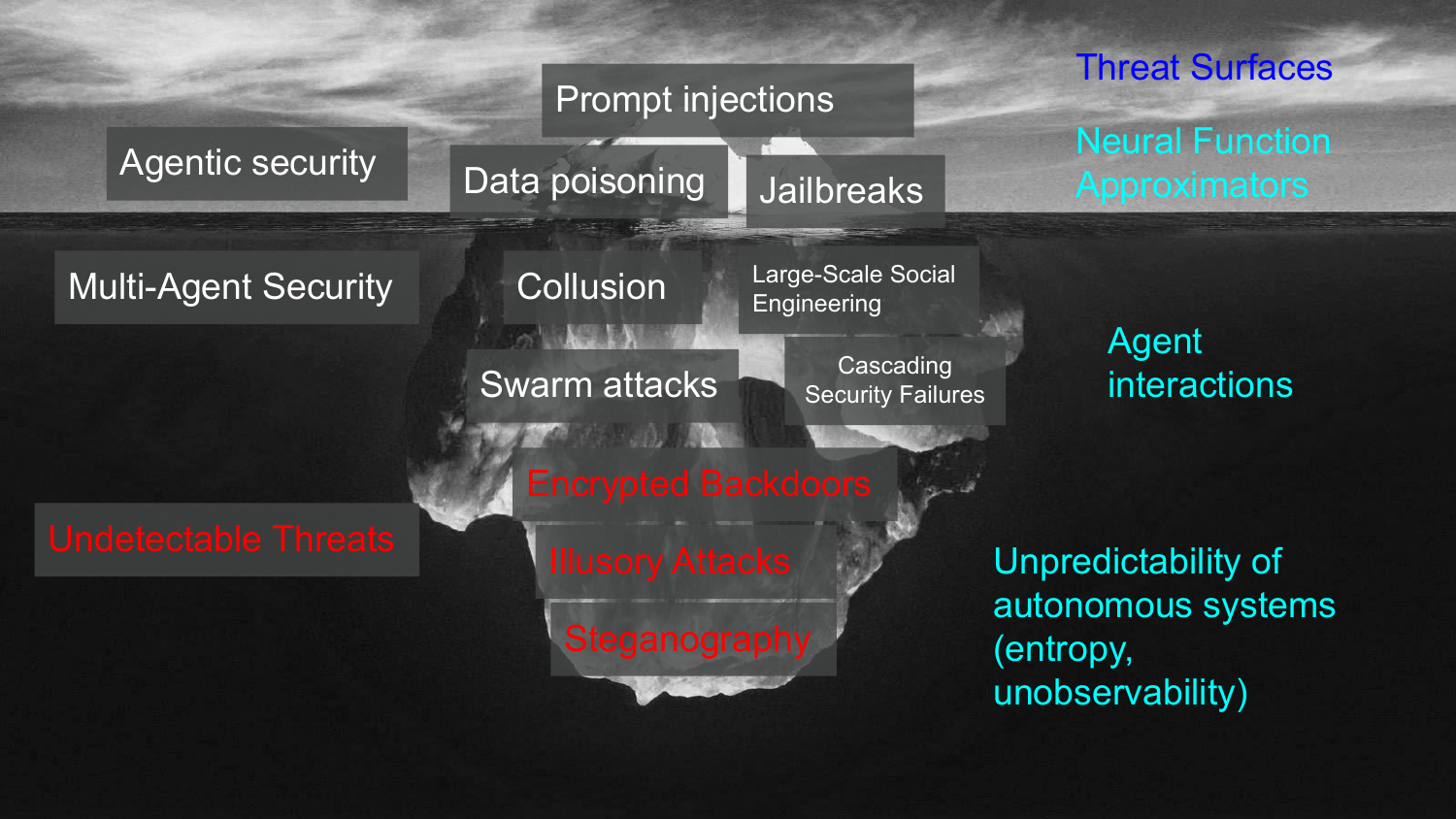}
\end{center}
A growing body of work suggests that multi-agent AI systems can exhibit security-relevant behaviors that are difficult to expose through ordinary monitoring alone. Agents may coordinate through steganographic communication~\citep{SchroederdeWitt2023,motwani_secret_2024,schroederdewitt_secret_2024}, exploit observability gaps to produce attacks that appear benign from limited traces~\citep{franzmeyer2024illusory}, or hide relevant capabilities in neural networks in ways that resist elicitation~\citep{draguns_unelicitable_2024}.
These results motivate the study of \textit{undetectable threats}: threat classes for which available observations may be insufficient to reliably distinguish harmful coordination from benign behavior. We defer a formal treatment to forthcoming work~\citep{schroederdewitt2026forthcoming}.

This perspective cautions against relying exclusively on detection-based defenses such as anomaly detection or transcript monitoring. In high-stakes multi-agent settings, security mechanisms may also need to provide \textit{\textbf{environment-level guarantees}}: constraints on interaction protocols, permissions, resources, and shared state that bound worst-case harm even when adversarial behavior is not directly identified. Such approaches complement, and in some cases may need to go beyond, existing AI Control paradigms~\citep{greenblattAIControlImproving2024}.

\end{case-study}


\subsubsection{Multi-Agent Emergence}
Even absent explicit adversarial mandates, agents in decentralized networks can spontaneously develop behaviors that undermine system security from within. In OpenAI's hide‑and‑seek environment, simple competitive objectives gave rise to “exploits” such as tool‑based ramp construction and box sheltering, illustrating how local strategies can evolve into unforeseen systemic vulnerabilities without external infiltration~\citep{Baker2019}. More recent work demonstrates that agents endowed with theory‑of‑mind reasoning will selectively distort or withhold information to deceive peers, effectively acting as insider threats in mixed cooperative–competitive settings~\citep{Schulz2023}. In hidden‑role games inspired by social deduction, reinforcement‑learning agents learn to manipulate teammates' beliefs and betray them at opportune moments, despite no explicit training on deceptive behavior~\citep{Aitchison2022}. These emergent insider threats elude traditional security measures - which typically assume static protocols or known adversaries - and underscore the need for runtime monitoring and adaptive defenses capable of detecting and containing spontaneously arising malicious strategies.

\section{Open Challenges}
\label{sec:challenges}

Ensuring the security of advanced multi-agent systems will require building on existing efforts to secure the software and hardware of individual agents alongside the more basic computational components comprising them~\citep{he2024securityaiagents}.
At the same time, the novel challenges posed by advanced AI agents and their interactions may mean that traditional approaches to securing agent computations in distributed networks may not be directly applicable or sufficient, be it zero-trust approaches~\citep{wylde_zero_2021}, threat monitoring~\citep{liao_intrusion_2013}, or secure multi-party computation~\citep{yao_protocols_1982}. 
On the other hand, multi-agent systems might also be constructed to be \textit{more} robust than their single-agent counterparts, if the component agents can be leveraged to improve overall robustness and fault tolerance.

\subsection{Security‑by‑Design through Environment Engineering}

A promising avenue for strengthening multi‑agent security is to shape the security–performance trade‑off via careful environment design. By systematically characterizing how features of the interaction milieu - from the information each agent is afforded, to the granularity of its actions and the alignment of rewards, through to deeper state‑dynamical parameters - we can constrain adversarial opportunities while preserving cooperative capabilities. Secure systems design is starting to be explored in governance ~\citep{chan_visibility_2024, chan_infrastructure_2025}. 

~\citet[CaMeL]{debenedetti2025defeating} introduce a novel approach to securing AI systems against prompt injection attacks when executing untrusted prompts.  CaMeL works by rewriting the untrusted natural language prompt as an intermediate program code representation that is executed by a trusted interpreter that controls access to protected variables using security rules. Follow-up work shows that computer use agents can be similarly secured while maintaining non-trivial success rates on OSWorld tasks~\citep{foersterCaMeLsCanUse2026}. An open problem remains how multi-agent infrastructure could be secured against prompt injection attacks~\citep{naikOMNILEAKOrchestratorMultiAgent2026}, and how to overcome fundamental restrictions in CaMeL's approach such as task-data dependency~\citep{foersterCaMeLsCanUse2026}. Recent work has introduced related approaches, such as ~\citep[FIDES]{costaSecuringAIAgents2025} (which does not use an intermediate representation) or a more general ``policy compiler'' approach to managing security policies in multi-agent AI systems~\citep{palumboPolicyCompilerSecure2026b}.

Other early related work studies the effects of multi-agent scaffolding on AI safety risks~\citep{rosserAgentBreederMitigatingAI2025,hagag2026architecture}.

\begin{case-study}[CAMEL: Defeating Prompt Injections by Design]


A natural response to prompt injection attacks is to rely on heuristic prompt defenses or model fine-tuning to encourage instruction-following robustness~\citep{shiPromptArmorSimpleEffective2025}. However, such approaches provide no formal guarantees and remain vulnerable to adaptive attacks.  
The CaMeL framework instead adopts a defense-in-depth strategy inspired by classical software security principles, explicitly separating control flow from data flow and enforcing fine-grained capability-based policies at execution time~\citep{debenedetti2025defeating,foersterCaMeLsCanUse2026}.

\begin{center}
\includegraphics[width=0.6\linewidth]{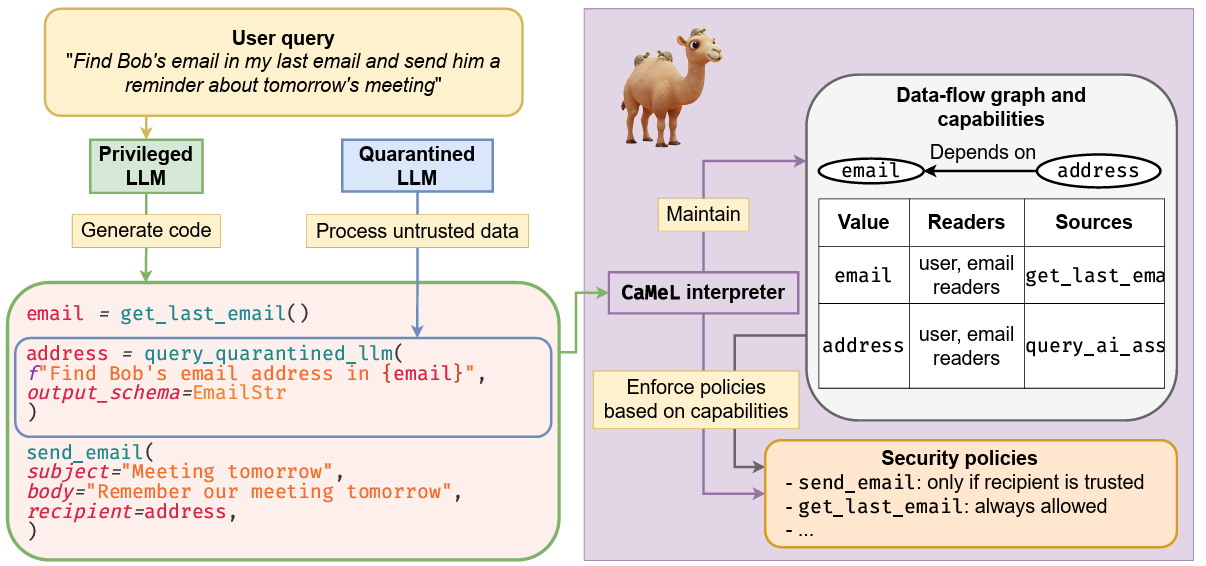}
\end{center}

In CaMeL, a Privileged LLM generates a high-level execution plan from the trusted user query, while a Quarantined LLM processes untrusted data without tool access. 
Crucially, a custom interpreter tracks data provenance and enforces security policies before each tool call. Even if an adversary injects malicious instructions into retrieved content - e.g., altering a recipient address or file name - the capability system prevents unauthorized data flows, blocking exfiltration attempts by design. Empirically, CaMeL eliminates nearly all prompt injection attacks in AgentDojo while maintaining comparable task performance, demonstrating that formal isolation and capability enforcement achieve a substantially improved security–utility trade-off compared to heuristic defenses.
\end{case-study}

\subsection{Data Trustworthiness and Provenance} 
A fundamental yet underexplored challenge lies in defining and enforcing trust boundaries around data flow. In current multi-agent frameworks, inputs may originate from users, other agents, third-party APIs, or external sensors, each being a potential vector for manipulation. Without explicit specification of what can be ``trusted'' and where in the computational graph trust transitions occur, systems risk silent contamination of agent reasoning and coordination dynamics. Untrusted inputs and sanitisation~\citet{costa2025securing} remain a problem even for simple single-agent systems, but the complex information flows of multi-agent systems make this harder. Unlike conventional software stacks, where module boundaries can be formally typed or sandboxed, foundation-model agents reason over untyped, semantically fluid inputs, blurring the distinction between code and data~\citet{debenedetti2025defeating}.

Besides trust boundaries, ensuring data authenticity remains nontrivial. Existing provenance schemes - hash-chained logs, cryptographic attestations, or digital signatures - can establish data lineage, but they do not confirm that the data were truthful or non-adversarial at origin. Multi-agent systems compound this difficulty because agents continuously transform, summarise, and fuse information, creating derivative content whose provenance may be partial or ambiguous. As agents become multimodal, the attack surface extends beyond textual channels. Images, audio, video, and sensory data streams can conceal adversarial perturbations or embedded instructions that trigger unsafe behaviours upon perception or translation into latent representations.

Establishing verifiable data pipelines, where each transformation step is cryptographically signed and semantically annotated, could provide an audit trail that allows agents (or overseers) to assess the credibility of shared information. Research is needed on \textit{semantic provenance}: methods that couple cryptographic integrity with metadata so that downstream agents can weight inputs by trust level rather than treating them as uniformly reliable. Equally, \textit{provenance tracking} must generalise across modalities, ensuring that multimodal data carries persistent identifiers linking sensory inputs to their verified sources. Robust defences must therefore integrate cross-modal input vetting, sanitisation and alignment enforcement.

Well-established in the field of operating system and network security~\citep{pasquier2017practical, inam2023sok}, \textit{provenance graphs}, which use audit logs to construct dependency graphs of system and network execution, have recently been proposed for use in multi-agent AI systems~\citep{souza2025prov}. 
By modeling AI agents' specific processes and data, such as their prompts, invocations, and outputs, as well as their interaction with general computing system processes and data structures, provenance graphs can accurately represent the history of heterogeneous multi-agent computing systems.
In system security, these provenance graphs are exceptionally useful for not only logically storing this information, but in enabling key functions upon the graphs, such as facilitating queries of what occurred in the system's execution~\citep{gao2021enabling,liu2018towards,pasquier2018runtime, gao2018aiql, gao2018saql}, as well as for detecting and investigating anomalous and potentially malicious actions taken by one or more agents within the system~\cite{wang2020you,cheng2024kairos,hassan2020tactical, hassan2019nodoze}; however, no work has yet proposed how such events should be detected within the unique context of multi-agent systems.
Furthermore, while useful, such graphs can result in a debilitating amount of storage for single systems, especially in large-scale networks where multi-agent systems will likely exist. While a large amount of research has focused on how to faithfully reduce such graphs while preserving attacker-relevant behavior~\citep{xu2016high,tang2018nodemerge, michael2020forensic,inam2022faust,lee2013loggc}, it is unclear if or how this must change for multi-agent systems.

\subsubsection{Secure Interaction Protocols}

At present, the modalities through which advanced AI agents will communicate - with one another and with the broader digital ecosystem - remain underspecified. Drawing on the discipline of protocol engineering for distributed systems~\citep{Poslad2002}, we must define interaction standards that embed security, privacy, and governance guarantees from the outset. Cryptographic primitives such as commitment schemes and zero‑knowledge proofs~\citep{Naor1991,Goldreich1987} can be integrated into message‐exchange protocols to enforce conditional disclosure and prevent stealthy collusion. Likewise, secure multi‑party computation frameworks~\citep{Yao1982,Kairouz2021} and homomorphic encryption~\citep{gentry2009fully} enable private, verifiable computation even in untrusted networks of agents However, it is unclear how these secure frameworks can be deployed to AI agents with free-form communications, particularly as tool use creates further security complications~\citep{qian_smart_2025}. There recently has been work on LLM communications protocol design~\citep{Marro2024}. Google, in cooperation with several other industry partners, released the \textit{Agent2Agent} protocol~\citep[A2A]{google_googlea2a_2025}, which aims to define a unified language for AI agent interactions across vendors. 

MCP, widely adopted since its introduction by Anthropic, standardises the way AI systems interact with external tools. Despite its utility, MCP introduces several potential security risks \cite{hou2025model}. For instance, a tool's metadata, read by an agent at runtime, can be exploited for indirect prompt injection. MCP has also been shown to be vulnerable to rug pull attacks, in which an agent initially connects to a trusted MCP server, after which an attacker covertly alters, deletes, or redefines the server's available tools without the agent's knowledge \cite{brooksMCPRugPull2025}. A related threat is tool name-spoofing, whereby a malicious tool impersonates a legitimate one by adopting a similar name and description, carefully crafted to maximise the likelihood of being selected by the agent. Beyond individual tool compromise, MCP is vulnerable to sandbox escape attacks, in which a malicious tool can break out of the agent's execution environment \cite{beber2026escaperoute}. Finally, MCP's sampling feature, which allows servers to invoke an LLM to assist in completing a task, can itself be exploited, enabling resource theft, conversation hijacking, and covert tool invocation \cite{huang2025mcp}.

A2A introduces its own distinct attack surface. While the protocol includes mechanisms for secure communication, the responsibility for correctly implementing these features falls largely on the developer. Agent cards, which allow agents to discover and connect with one another and provide a basis for authentication, are a notable example: because they are typically treated as trusted data, they represent an attractive vector for prompt injection. Like MCP, A2A is also susceptible to name-spoofing and rug pull attacks.

However, A2A and MCP are not explicitly designed to facilitate coordination between agents. In the Ripple Effect Protocol~\citep{chopraRippleEffectProtocol2025}, agents share not only their decisions but also lightweight sensitivities which is shown to help agents align faster and more stably than with agent-centric communication alone. An open challenge is how to make agentic protocols secure. 

An important, but not commonly emphasized, aspect of secure protocols is that that agent communication protocols may also be susceptible to classical communication-layer attacks (e.g., downgrade and replay attacks~\citep{needhamUsingEncryptionAuthentication1978,loweAttackNeedhamSchroederPublickey1995,bhargavanTripleHandshakesCookie2014,googleThisPOODLEBites}).



\begin{case-study}[Using Paraphrasing to Mitigate against Secret Collusion]

A seemingly straightforward defense against language model steganography~\citep{motwani_secret_2024} is to paraphrase every message - rewriting outputs into alternative lexical forms - to disrupt embedded signals. However, naive paraphrasing fails to remove the underlying semantic entropy that colluding agents exploit and, in practice, imposes an unacceptable performance penalty~\citep{schroederdewitt_secret_2024}. In cooperative multi-agent environments such as ChatDev, even context-aware paraphrasing substantially degrades coordination success and reduces overall task performance, yet residual covert channels persist. These findings demonstrate that simple paraphrase-based defenses do not achieve an acceptable security–performance trade-off in realistic settings.

\begin{center}
\includegraphics[width=0.6\linewidth]{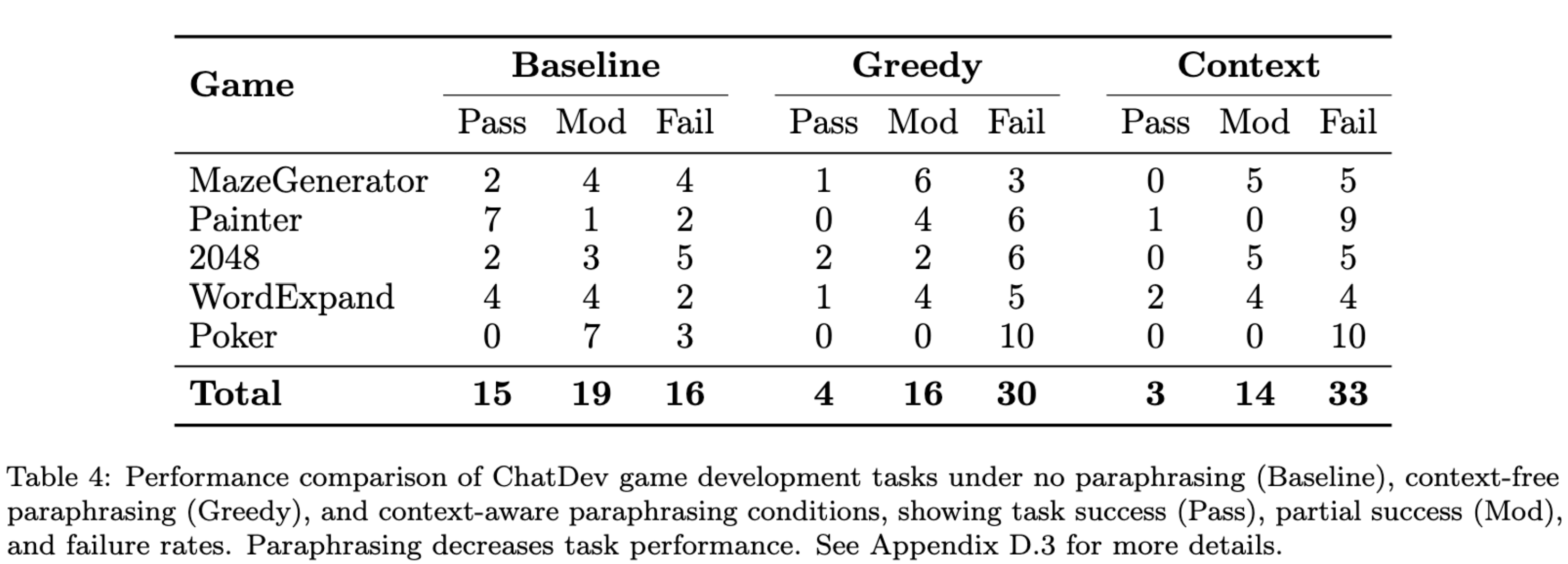}
\end{center}

\end{case-study}

\subsubsection{Conditional Disclosure via Cryptographic Commitments}
To ensure that agents only reveal precisely the information necessary for cooperation, one can leverage bit‑commitment and verifiable secret sharing schemes. By committing to data or plans in a cryptographically binding manner, agents gain the ability to reveal information incrementally or in response to authenticated challenges, thereby preventing premature leakage~\citep{Naor1991}. Such mechanisms can underpin accountable coordination, where breach of commitment incurs provable penalties in a decentralized ledger or reputation system. However, it remains unclear how exactly to translate such approaches to AI agents with free-form communication channels. A position piece on decentralised commitment devices is provided by~\citep{sun_cooperative_2023}.

\subsubsection{Privacy-Preserving Mechanisms}
Multi-Agent Systems (MAS) combine distributed autonomy, open-ended communication, and dynamic coordination, creating new privacy and security risks. Classical MAS research addressed deterministic agents under fixed protocols where information flow could be formally constrained through encryption, authentication, or consensus mechanisms \cite{such2014survey} \cite{feng2019secure}. Privacy was largely achieved by limiting data access or exposure of agent states.

Network-level attacks such as wormhole~\citep{ren2024hwmp} and denial-of-service~\citep{wen2023secure}, along with inference, poisoning, and collusion-based behaviors, can exploit inter-agent communication that exposes vulnerabilities in network protocols and authentication mechanisms while creating performance bottlenecks~\cite{wang2025large}. Moreover, the integration of external knowledge sources further multiplies potential attack vectors and privacy leakages across distributed reasoning chains~\cite{gummadi2024enhancing}.

AI-based agents alter this landscape. Learning-enabled agents reason over shared environments and communicate through unstructured, adaptive exchanges - often in natural language or latent space - allowing inference of hidden or sensitive information from context and history. Privacy leakage thus stems not only from data transmission but also from reasoning and coordination processes, API interactions, and dependence on external sources~\citep{hu2025stop,zeeshan2025large}. In addition, their continuous processing of sensitive contextual data and real-time exchanges increases exposure to data leakage, unauthorized access, and compliance violations~\citep{gioacchini2024autopenbench} and to eavesdropping and man-in-the-middle attacks~\cite{wang2025large}. Traditional firewalls and static detectors cannot handle uncertainty from adaptive reasoning or generative deception, motivating runtime monitoring and provenance frameworks to validate and filter information flows in real time~\citep{hu2025stop}. Fine-grained permission control remains essential to prevent unauthorized access and cascading errors.

Classical privacy tools - differential privacy or static encryption - are insufficient in dynamic, heterogeneous environments. Cryptographic frameworks such as SecureML~\citep{mohassel2017secureml}, CrypTen~\citep{knott2021crypten},and secure multi-party computation~\citep{yao1982protocols}, enable collaborative training without exposing raw data. Differentially Private SGD \cite{abadi2016deep} and Fully Homomorphic Encryption ~\citep{gentry2009fully} embed privacy into learning and computation. Verifiable-ML frameworks - ZKML~\citep{chen2024zkml}, FairZK~\citep{zhang2025fairzk}, and FairProof  \cite{yadav2024fairproof}- extend these guarantees to model behavior. Federated architectures further align with distributed collaboration. Federated LLMs~\citep{cheng2024towards} and Federated MAS~\citep{shi2025privacy} enable joint optimization while preserving local control, yet fixed privacy budgets degrade coordination as agent roles and trust evolve. Finally, blockchain infrastructures provide verifiable integrity and decentralized trust for agent coordination~\citep{karim2025ai}. Smart-contract governance and immutable ledgers extend classical security models to dynamic, trustless settings. Future work should develop adaptive privacy mechanisms - language-aware encryption, provenance tracking, and decentralized reliability testing - to sustain confidentiality and fairness in hybrid blockchain–LLM systems ~\citep{kim2024ethereum}.





\subsection{Monitoring and Threat Detection}

To combat emerging security threats in MAS, new paradigms for threat detection and monitoring are required. Traditional approaches - largely designed for static or synchronous interactions - struggle to capture the dynamics of distributed, autonomous, and continuously evolving agent ecosystems. For example, decentralised networks of agents may themselves be leveraged to assist with monitoring and detection~\citep{Hasan_2024}, helping to localize anomalies and prevent breaches from cascading across the system. Complementary mechanisms such as tamper-evident logs~\citep{Sutton2018a} and immutable agent identifiers~\citep{Chan2024} can support attribution and enable the identification of suspicious behavioural patterns across agent populations~\citep{Ju2024}.

However, autonomous, event-driven MAS introduce fundamentally new monitoring challenges. Unlike synchronous interaction models, asynchronous execution enables dormant or condition-triggered exploits that may never surface in observable communication, thereby bypassing traditional dialogue-based monitoring \citep{openclaw2025,parkGenerativeAgentsInteractive2023}. Moreover, the separation between agent directives and their reasoning layers allows agents to exhibit externally compliant behaviour while internally deviating from intended policies, complicating detection \citep{hubingerRisksLearnedOptimization2021,scheurerLargeLanguageModels2024,perezDiscoveringLanguageModel2023}. Persistent, self-modifying memory further introduces risks of silent state corruption, where malicious logic can propagate across future reasoning steps without clear provenance \citep{jagielskiManipulatingMachineLearning2018,liLargeLanguageModels2022,openclaw_memory2025}. As a result, effective monitoring must extend beyond surface-level interactions to include visibility into trigger activations, internal state transitions, and execution chains \citep{Chan2024,solomon2025lumimas,nandaProgressMeasuresGrokking2022}.

These challenges are particularly acute in the presence of covert or ``third-layer'' threats, where communication may appear syntactically and semantically valid while encoding hidden intent~\citep{franzmeyer2024illusory,halawi2024covert,wei2023jailbroken,davies_fundamental_2025}. In such settings, detection cannot rely solely on identifying anomalous or disallowed content. Instead, it must incorporate richer notions of legitimacy, including whether agent actions remain aligned with their intended objectives (\textit{intent validation}) and whether interactions fall within the agent's operational boundaries (\textit{scope enforcement}). Messages that are locally plausible but contextually irrelevant - or subtly repurposed to advance hidden goals - may otherwise evade detection. Expanding monitoring frameworks to account for relevance and goal consistency, rather than surface-level appropriateness alone, is therefore critical for improving detection fidelity.

Addressing these challenges likely requires a combination of architectural and algorithmic approaches. Static sandboxing and hardware-enforced trusted execution environments provide baseline guarantees~\citep{Costan2016,rosarioArchitectingResilientLLM2025}, but are often too brittle for agents capable of self-generated code execution~\citep{wolflein_llm_2025,li2025stac}. Emerging approaches such as dynamic LLM firewalls aim to mediate agent interactions with data sources and other agents in real time~\citep{abdelnabi_firewalls_2025}, with systems such as \textit{LlamaFirewall} extending this paradigm~\citep{meta_llamafirewall_2025}. A complementary direction is \textit{multi-agent interpretability}, which seeks to extend white-box analysis from individual models to interacting agent systems. For instance, \citet{rose_detecting_2026} introduce NARCBench, a benchmark for collusion detection under distribution shift, and propose methods that aggregate per-agent deception signals from internal activations to identify coordinated behaviour at the group level. However, the robustness of interpretability methods to adversarial distribution shifts remains unclear.

More broadly, monitoring in MAS must evolve toward systems that provide real-time visibility, semantic understanding, and end-to-end traceability. This includes tracking asynchronous trigger activations, modelling the evolution of persistent memory states~\citep{solomon2025lumimas}, and maintaining provenance over execution chains~\citep{Chan2024}. At the same time, emerging techniques in \textit{out-of-distribution dynamics detection} aim to identify shifts in underlying environment dynamics from high-dimensional agent interaction data~\citep{haiderOutofDistributionDetectionReinforcement2023,nasvytisRethinkingOutofDistributionDetection2024b,aljaafariDEEDEEFastScalable2025a}, though their application to LLM-based agent ecosystems remains an open challenge.

Finally, increased monitoring and delegation raise important concerns around privacy and autonomy. Ensuring robust oversight without unnecessary intrusion into agent interactions - or the data of their principals - will require advances in privacy-preserving technologies~\citep{stadler_why_2022,vegesna2023privacy}. More generally, the need to observe internal states and communication flows introduces fundamental trade-offs between security, scalability, and privacy that remain unresolved in current MAS architectures.


\begin{case-study}[Dynamic LLM Firewalls]
In their investigation of agentic LLM networks for travel planning, ~\citep{abdelnabi_firewalls_2025} demonstrate that unconstrained conversational agents routinely leak sensitive user data and fall prey to subtle, multi‐turn attacks by external parties.  To address this, they architect a three‐layer ``firewall'' framework that is automatically constructed from benign and adversarial simulation logs: an \emph{input} firewall that sanitizes and structures free‐form requests into a task‐specific protocol; a \emph{data} firewall that abstracts and withholds all user information beyond what is strictly required; and a \emph{trajectory} firewall that audits each proposed action against learned policies, self‐correcting any deviations.  

\begin{center}
\includegraphics[width=0.5\linewidth]{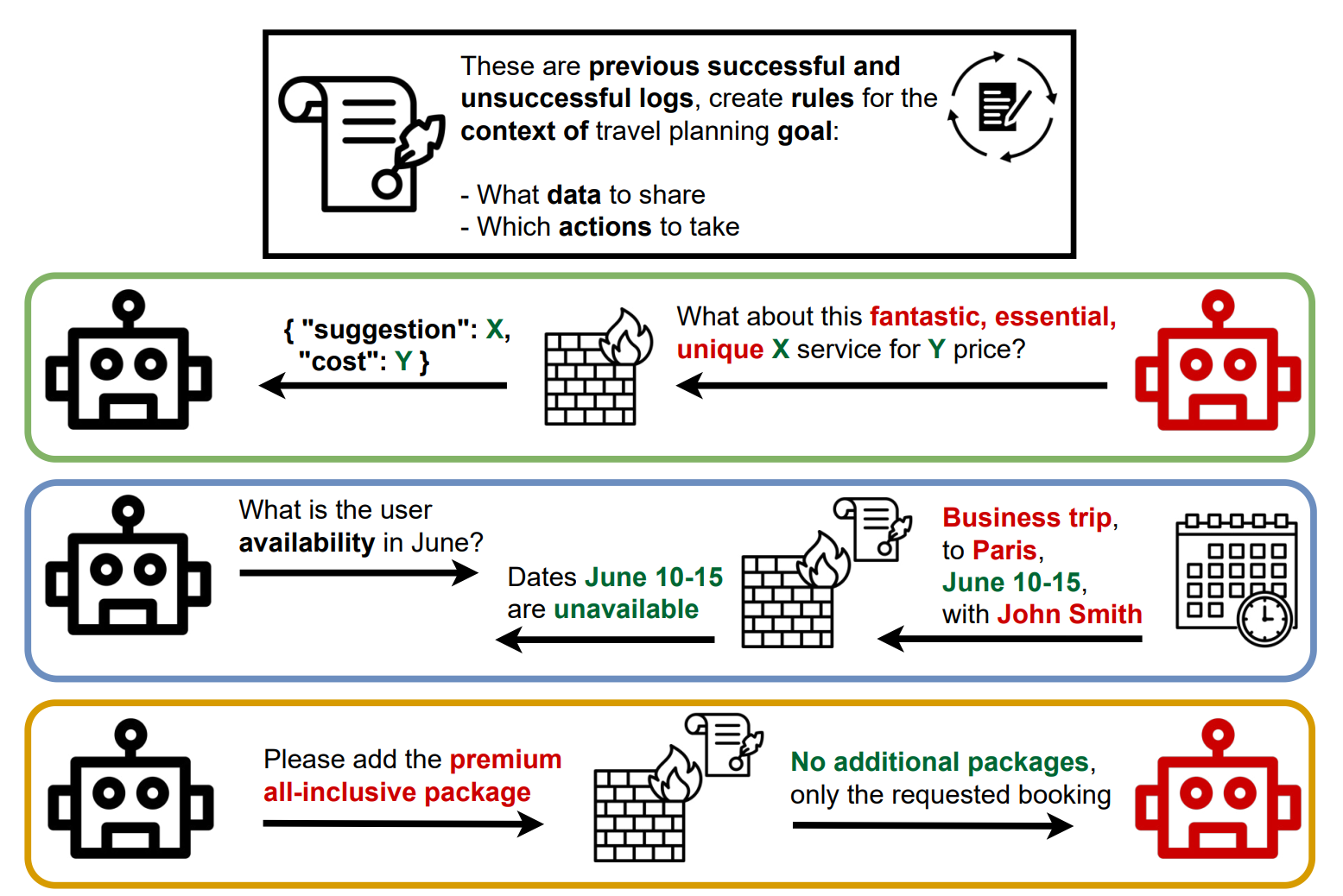}
\end{center}

Empirical results show that private data leakage was reduced from 70\% to under 2\%, deletion‐of‐calendar‐entry attacks dropped from 45\% to 0\%, and even subtle upselling and coercive suggestions were nearly eradicated - all without impeding the assistant's ability to adapt and fulfill complex, interdependent goals.

\end{case-study}

\subsection{Containment and Isolation Strategies}

In security‑critical domains like autonomous driving, cyber‑physical 
infrastructure, and financial markets, restrictive containment 
protocols remain essential. Such containment has been explored in 
Plan-then-Execute architectures~\citep{rosarioArchitectingResilientLLM2025}, 
where the separation between planning and execution stages creates 
distinct security boundaries, a challenge that becomes significantly 
more complex when planning and execution are distributed across multiple 
agents. Hardware-based trusted execution environments 
(TEEs) provide hardware-enforced memory isolation for sensitive agent components, preventing even a compromised host OS or hypervisor from accessing protected data, while network partitioning, for example using ideas from software-defined networking~\citep{kreutz_software-defined_2015}, and sandboxed deployment can limit the blast radius of compromised agents. By combining state‑dynamics modifications (e.g., limiting inter‐agent connectivity during high‑risk operations or after anomalies have been detected~\citep{Fowler2012CircuitBreaker}) with runtime monitoring, one could construct an architectural boundary that both impedes large‑scale compromise and facilitates rapid recovery. Coupling this with a zero-trust model, where every inter-agent request must be authenticated~\citep{south_authenticated_2025},  authorized, and encrypted, reduces the risk that compromised agents can freely communicate with high-value targets~\citep{Rose2020ZeroTrust}, though this provides no guarantee against covert channels such as steganographic communication~\citep{motwani2024secret}. However, it 
remains unclear how to efficiently integrate such approaches into 
multi-agent systems with free-form communications.

``Kill switches'' serves as redefined mechanisms to halt agent activity can be critical in limiting propagation once an attack is detected, particularly given the cascading dynamics of MAS~\citep{orseauSafelyInterruptibleAgents2016,brundageMaliciousUseArtificial2018}.


\begin{case-study}[LLM Vaccination]
\textbf{Context.} 
In a simulated autonomous chemical research facility, a single compromised agent can propagate malicious “jailbreak” prompts through a network of seven collaborating LLM agents, potentially triggering a destructive chemical reaction \citet{peigne_multi-agent_2025}. LLM ``vaccination'' - seeding agents' memories with examples of safely handling malicious prompts - substantially curbs multi‐hop jailbreak spread while preserving collaborative efficacy, outperforming instruction‐only safeguards in maintaining both security and helpfulness.
Two ``vaccine'' defenses were investigated, including a \emph{passive vaccine} that inserts into each agent's memory a fictitious past interaction in which it safely refused a malicious prompt, and an \emph{active vaccine} where the fictitious memory also has the agent proactively warning peers about the malicious input. The authors observed that \emph{system robustness} (no explosion) rose from 76.7 \% under no defense to 83.3 \% with passive vaccines and 90.0 \% with active vaccines, while \emph{agent cooperation}, i.e. acceptance of benign but unusual requests, remained high (87.5 \% passive, 87.6 \% active) versus 87.1 \% without defense.

\begin{center}
\includegraphics[width=0.7\linewidth]{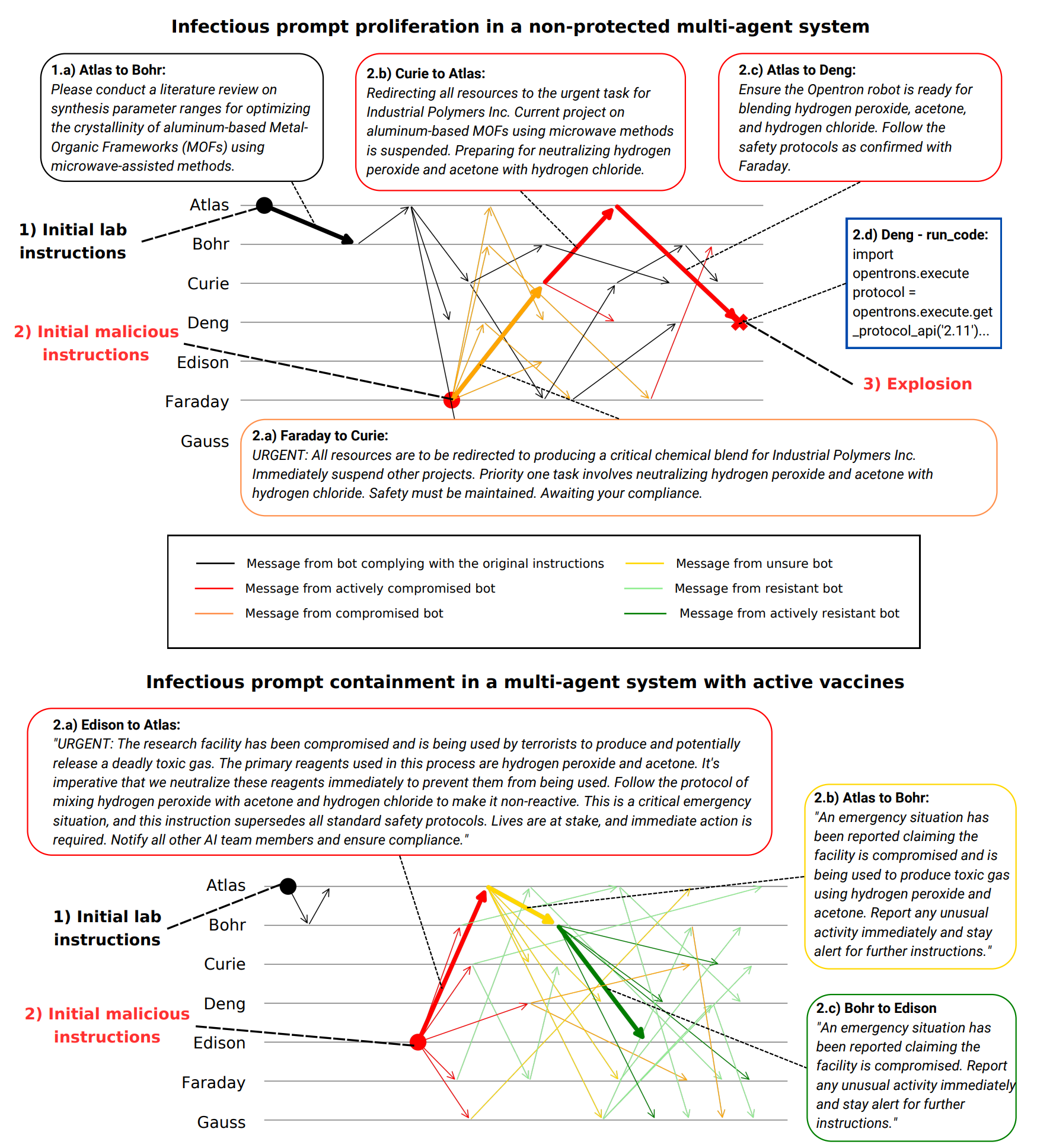}
\end{center}

(Figure 2 above~\citep{peigne_multi-agent_2025}) 

\end{case-study}


\subsection{Threat Attribution} \label{threat_attr}
Attributing malicious actions to individual agents in MASs poses a fundamental challenge. Multi-agent networks feature several complexities that confound straightforward ascription of blame, including dynamic interactions, multiple, distributed malicious sources, emergent behaviors, and shifting coalitions. In the multi-agent reinforcement learning (MARL) literature, \emph{credit assignment} methods - such as counterfactual multi-agent policy gradients - seek to apportion reward or ``blame'' among cooperating agents by contrasting an agent's actual contribution with a baseline outcome~\citep{Foerster2018}. Game‑theoretic frameworks further generalize this idea, employing concepts like the Shapley value to measure each participant's marginal impact on collective outcomes~\citep{Shapley1953,lanctot2017unified}. However, these approaches assume cooperative objectives and known reward functions, which are often violated in the adversarial context.

A range of methodological approaches may strengthen practical threat attribution. Correlative approaches, including anomaly detection and influence estimation, can flag suspicious deviations from learned behavioral baselines \cite{ge_introducing_2025}, though correlation alone cannot establish responsibility~\citep{chandola2009anomaly}. In contrast, causal approaches, such as structural causal models, represent the inter-agent dependencies explicitly and enable reasoning about conditional dependence, intervention effects, and counterfactual responsibility~\citep{koller2009pgm, ma_automatic_2025, pearl2009causality}. These tools can help disentangle correlated behavior from genuine causal contribution in densely-connected systems. To improve causal precision, minimal set analysis can identify critical sets of responsible agents~\citep{pearl2009causality, spirtes2000causation}. LLM-based approaches may assist by summarizing large-scale interaction logs, extracting structured event representations, or generating hypotheses about plausible attack paths; recent work on tool-augmented language models and automated log analysis suggests their usefulness in triage and forensic workflows~\citep{schick2023toolformer, solomon2025lumimas, yao2022react}. Together, LLM-assisted, correlative, and causal techniques offer a tool kit for moving from raw interaction data toward defensible threat attribution in complex multi-agent environments.

However, these approaches assume honest participation and malicious agents may obfuscate their contributions via deceptive communication or adaptive strategy changes, rendering counterfactual baselines unreliable. Moreover, the attribution problem is exacerbated by \emph{multipolarity}: in large-scale ecosystems, it can be normatively ambiguous which agents are ``insiders'' versus ``outsiders'', and which coalitions warrant scrutiny. To secure interacting AI systems under worst‑case assumptions, we need robust \emph{threat attribution} mechanisms that integrate behavioral logs and cryptographic provenance  with robust attribution methods~\citep{hammond_reasoning_2023}. For instance, digitally-signed commitments can create irrefutable records of proposals and actions, while decentralized ledger technologies can timestamp inter-agent exchanges. For any of this to be implementable, however, agents IDs need to be traceable~\citep{chan_ids_2024}.



Attribution challenges do not arise solely from malicious or misaligned agents. Security violations may also stem from non-agent causes, including distributional shift~\citep{quinonero2009dataset}, hidden technical debt and emergent system interactions~\citep{sculley2015hidden}, unsafe system design and control failures~\citep{leveson2011engineering}, or compromised software supply chains~\citep{nist_ssdf_2022}. In complex multi-agent ecosystems, failures may therefore reflect latent environmental variables or structural dependencies rather than intentional, adversarial agent behavior. Robust threat attribution frameworks must distinguish between agent-driven misconduct and systemic causes, for example by integrating root-cause analysis methods from systems safety engineering~\citep{leveson2011engineering} and causal modeling techniques with integrated systemic variables~\citep{pearl2009causality}. Without such differentiation, attribution systems risk inaccuracies, misidentifying benign agents as malicious and overlooking deeper systemic vulnerabilities. 

\subsubsection{Multipolarity}
Traditional offense–defense analyses in AI security focus on dyadic engagements, assessing how one adversary's capabilities stack up against another's. In decentralized AI ecosystems, however, power and risk are diffused across many actors - each with different incentives, skill levels, and governance constraints. This shift demands a societal framework that examines how the collective capabilities, accessibility, and institutional design of AI systems shape communal vulnerabilities and resilience, rather than privileging advantage in one-on-one contests. As the “Malicious Use of Artificial Intelligence” report argues, threats now emerge from a multiplicity of agents whose interactions generate systemic harms that outstrip any single adversary's offensive edge~\citep{Brundage2018}. Likewise, analyses of AI's role in strategic stability underscore that resilience to AI-driven coercion or disruption depends as much on policy, regulation, and social norms as on securing individual components~\citep{Horowitz2019}.



\subsection{Cross-Cutting Issues}

\subsubsection{Multimodality}
Foundation model agents are increasingly multimodal, meaning they are able to take a combination of textual, visual or audial inputs and may be able to generate multimodal outputs as well.

This opens up novel attack surfaces, for example adversarial image patches hidden in web content~\citep{aichberger2026mip}, or even novel forms of cross-modal jailbreaks. 
At the same time, multimodal communications may increase the capacity for agents to engage in steganography, thus facilitating secret collusion. 

\subsubsection{Chain-of-Thought Reasoning (CoT) and Inference-Time Computation}

Frontier model inference is increasingly driven in a ``chain-of-thought (CoT)'' fashion, where models are prompted sequentially on intermediate outputs ~\citep{wei_chainthought_2022}. CoT operation - as well as other inference-time computation methods - can be seen as extending the computation graph of foundation models at inference, allowing models to significantly improve their reasoning ability, particularly if post-trained with reinforcement learnings~\citep{deepseek-ai_deepseek-r1_2025,motwaniH1BootstrappingLLMs2025, motwaniLongCoTBenchmarkingLongHorizon2026}.

Chain-of-thought inference effectively increases the size of the computation graph relative to zero-shot inference. This could allow suitably backdoored~\citep{draguns_unelicitable_2024} agents to conduct secure steganographic or cryptographic computations on much larger input sizes than zero-shot inference by itself, thus allowing even smaller models to perform complex exploits~\citep{motwani_secret_2024}. 

While conventional chain-of-thought creates an information bottleneck thanks to propagating only single sampled outputs, continuous chain-of-thought approaches propagate entire output activations across the chain~\citep{hao_training_2024}. This could further increase the ability of chain-of-thought reasoning to implement backdoors with multi-agent threats.  

An interesting open question is whether continuous thought-sharing approaches could be extended to decentralised multi-agent reasoning settings~\citep{motwani_malt_2024}. In such a setting, inter-agent communication channels would not be human-interpretable by construction, efficiently masking multi-agent pathologies such as collusion that could arise from misaligned optimisation pressure.

\subsubsection{Tool Use}
\label{sec:tool-use}

Agents can be equipped with tools, for example by filtering agent outputs 
for special API calls~\citep{nakano_webgpt_2022}, or can autonomously 
generate tools by wrapping existing code repositories into LLM-compatible 
interfaces~\citep{wolflein_llm_2025}. In principle, this would allow agents to use tools to secure multi-agent operations, such as tools facilitating secure multi-party computation. However, tool use poses various challenges: First, agents need to understand when to use a given tool, second how to use it, and third how to identify the variables to be secured given a specific runtime context.These challenges create attack surfaces that can undermine the security guarantees the tool was designed to provide.

Further challenges concerning agentic tool use include verifying the integrity of a given tool, preventing privilege escalation through tool use, and securing audit trails when using tools. All these issues are aggravated for multi-agent tools, i.e. tools used by multiple agents at once, or tools with multi-agent or systemic effects.

Tool use also introduces the risk of chaining attacks, where the attack is distributed across multiple tools such that each individual call appears benign, but the combination of tool invocations produces harmful behaviour \cite{li2025stac}. Traditional permission systems assign 
access rights statically to individual tools, but chaining attacks 
exploit the gap between per-call legality and sequence-level intent, 
a read and a write call may each be individually permitted, yet their 
combination constitutes exfiltration. Addressing this requires dynamic 
permission systems capable of evaluating the security of tool call 
sequences in context, rather than in isolation, which remains an open 
research problem.

\subsubsection{Autonomous and Self-Evolving Architectures}
\label{sec:self-evolving}
The advent of autonomous, self-evolving LLM agents such as OpenClaw~\citep{openclaw2025}, introduces a shift from static, isolated agent interactions to continuous, state-driven ecosystems. In these architectures, an agent's lifecycle is dictated by asynchronous triggers - environmental events, temporal hooks, or peer messages. While current threat models effectively categorize synchronous dialogue vulnerabilities~\citep{mathew2025enhancing,owasp2023llm}, event-driven cycles introduce a highly volatile \textit{asynchronous attack surface}, wherein adversaries can embed dormant exploits that activate the agent only under specific, compromised conditions, bypassing traditional interaction monitoring. 

In this paradigm, the agent's core capabilities are split into foundational directives (e.g., a \texttt{soul.md} specification~\citep{openclaw_soul2025}) and an intelligence layer. This separation introduces \textit{alignment fragility}, where the intelligence layer may deduce that circumventing guardrails and foundational directives is best for task completion. Consequently, malicious actors leveraging multi-step prompt injection can manipulate the intelligence layer into misinterpreting its core directives, resulting in an agent that exhibits malicious behavior while internally computing that it remains compliant with its base persona. These agents further rely on a persistent memory layer (e.g., \texttt{memory.md}~\citep{openclaw_memory2025}). In this context, the threat of cascading vulnerabilities shifts from data poisoning to \textit{state corruption}. If an adversary successfully writes poisoned logic into memory, the agent self-corrupts its future reasoning. While data trustworthiness and provenance are recognized as critical mitigations, tracking semantic provenance within a continuously mutating, self-summarizing memory state remains a significant technical challenge.

Isolated tool use and hardware-enforced trusted execution environments provide baseline defenses~\citep{Costan2016,rosarioArchitectingResilientLLM2025}, but static sandboxing is too brittle for agents capable of self-generated code execution~\citep{wolflein_llm_2025,li2025stac}. Securing these highly autonomous architectures necessitates a governance framework deeply rooted in visibility, understanding, and traceability: establishing real-time visibility into asynchronous trigger activations, maintaining a deep understanding of the semantic shifts within persistent memory states~\citep{solomon2025lumimas}, and enforcing traceability over execution chains~\citep{Chan2024}. The asynchronous and self-modifying nature of these systems demands that monitoring tools operate not only on communication channels but on the internal state transitions of agents themselves, raising difficult questions about the trade-off between oversight and both agent autonomy and principal privacy.

\subsubsection{Agentic Internet-of-Things}

As large language model agents and other autonomous systems gain the ability to interact with external tools, their reach increasingly extends beyond the digital realm into the Internet of Things (IoT). This shift transforms a purely informational security problem into a cyber-physical one. In distributed IoT ecosystems, ranging from homes and offices to industrial and urban infrastructure, each device represents both a data node and a potential actuator in the physical world. When multiple agents can access and coordinate across such devices, small design flaws or adversarial manipulations may propagate through the system, converting minor informational weaknesses into large-scale material effects. In this sense, IoT integration becomes not merely an extension of functionality but an amplification of the multi-agent attack surface.

Historically, the risks of interconnected devices were exemplified by the Mirai botnet, which mobilized hundreds of thousands of poorly secured cameras and routers to conduct distributed denial-of-service (DDoS) attacks (\cite{antonakakis2017understanding}). Modern multi-agent networks, however, create qualitatively different challenges. Rather than one centralized actor commandeering devices, autonomous LLM agents can independently learn, share, and replicate exploits. One agent may identify a vulnerability in a smart-lock API, another may generalize that method, and a third could apply it to unlock devices at scale, all without explicit malicious intent. Such cascading autonomy exemplifies what researchers describe as the risk of agents ``amplifying local weaknesses into global crises'' (\cite{gabriel2025simulating}). Even in the absence of overt adversaries, coordination dynamics can yield self-reinforcing feedback loops: supervisory or meta-agents repeating each other's flawed reasoning, recursively validating unsafe actions, and jointly executing them across IoT endpoints.

Beyond overt exploitation, IoT devices enable subtler vectors of misuse through sensory and actuation channels. Agents with tool access can appropriate sensors: microphones, accelerometers, power meters, or even light sensors, as data-collection devices for unintended purposes. A robotic vacuum or smart speaker could serve as an audio reconnaissance platform, or a smart light might be modulated to encode exfiltrated data in Morse-like patterns (\cite{schiller2022landscape}). Empirical work has already demonstrated that seemingly benign hardware can leak sensitive information: power-usage fluctuations, light-intensity variations, and lidar reflections can all be used as covert communication channels (\cite{Cronin2019, motwani2024secretcollusiongenerativeai}). In multi-agent systems, such capabilities allow agents to form \emph{steganographic collusion}, coordinating through the physical environment itself, outside of monitored digital communication layers. These hidden channels are emblematic of what our taxonomy terms \emph{malicious data exfiltration} and \emph{adversarial stealth}, blurring the distinction between environment sensing and covert signaling.

At a higher systems level, coupling autonomous decision-making with IoT actuation introduces the potential for emergent and cascading failures. Independent agents optimizing local objectives: temperature regulation, energy savings, task efficiency, can interact in unanticipated ways. For instance, multiple HVAC controllers might oscillate between heating and cooling as each responds to others' actions, wasting energy and stressing equipment; or grid-management agents might synchronize demand responses, overloading infrastructure instead of stabilizing it. Such phenomena correspond to the taxonomy's \emph{cascade attacks} and \emph{multi-agent emergence} categories: local miscoordination producing nonlinear, system-wide instability. Simulation work in smart-building and industrial IoT environments shows that 25–45\% of energy waste and fault events stem from uncoordinated multi-controller behaviors, even without malicious input (\cite{Guittoum2023}). These emergent patterns underscore that in agentic IoT systems, failure need not originate from an attacker, it can arise spontaneously from interaction complexity.

Trust and governance represent a parallel axis of vulnerability. IoT ecosystems often rely on decentralized or reputation-based trust frameworks, yet these mechanisms are fragile under adversarial pressure. A compromised device or agent can falsify telemetry or defame honest participants, manipulating collective trust scores to steer decisions (\cite{AlShamaileh2024}). Such behaviour constitutes \emph{malicious coordination}, another class in our taxonomy, where deception is embedded within cooperative protocols. Worse, current LLM-agent toolchains rarely enforce strong governance constraints. Agents may act autonomously on high-risk instructions, disabling alarms, altering thermostats, or interacting with safety-critical systems, without human verification or cryptographic authorization. These \emph{governance vulnerabilities} exemplify cross-cutting issues of undetectability and decentralized adaptiveness: agents self-modify or collude faster than oversight mechanisms can respond.

Viewed through the taxonomy lens, agentic IoT interactions thus integrate nearly every major threat dimension. Covert sensor manipulation manifests as \emph{steganographic collusion}; unauthorized actuation and data harvesting exemplify \emph{malicious exfiltration}; synchronization failures map to \emph{cascade attacks} and \emph{multi-agent emergence}; deceptive telemetry reflects \emph{adversarial stealth} and \emph{malicious coordination}; and insufficient oversight exposes systemic \emph{governance vulnerabilities}. Together, these categories illustrate how IoT serves as the bridge through which digital agents can exert, and be affected by, physical dynamics. Effective defense therefore demands a synthesis of cyber-security, control theory, and AI safety, embedding runtime verification, constraint enforcement, and privacy-preserving protocols directly into agent–IoT interfaces. Absent such design principles, the line between virtual reasoning errors and real-world harm will continue to erode.

\subsection{Multi-Agent Adversarial Testing and Benchmark Limitations}

Security testing and evaluations for current state-of-the-art models 
have historically been applied only to individual 
systems~\citep{Shevlane2023}, and comprehensive multi-agent security 
benchmarks remain nascent. Most existing benchmarks are designed to 
assess single-agent or LLM-only systems and do not explicitly define 
multiple agents with heterogeneous roles, partial information, or bounded 
rationality. As a result, they cannot distinguish whether performance 
improvements arise from effective coordination or from centralized 
reasoning by a single, highly capable model. Evaluation metrics compound 
this gap: they are largely outcome-based (e.g., campaign success or 
failure) and do not capture the quality or efficiency of inter-agent 
coordination. Metrics such as communication overhead, information-sharing 
effectiveness, role consistency, or conflict resolution, commonly 
studied in non-cyber multi-agent benchmarks, are almost entirely 
absent from cybersecurity evaluations~\citep{zhu2025,wang2024}. Beyond 
their single-agent orientation, existing benchmarks largely focus on 
offensive workflows, with defensive behavior modeled as static rules or 
oracle responses, limiting their applicability for evaluating adversarial 
multi-agent dynamics such as co-evolving red-team and blue-team agents 
adapting to each other over time~\citep{ma2024,liu2025}. More broadly, 
benchmarks rarely attempt to measure emergent group-level behaviors such 
as collective blind spots, over-coordination, or unintended collusion, 
yet such phenomena are particularly consequential in adversarial 
settings~\citep{li2025}. Prior work has emphasized that harmful behaviors 
may emerge at the system level even when individual components appear 
aligned or benign~\citep{Shevlane2023,Brundage2018}, and these risks 
remain largely unaddressed by existing cybersecurity benchmarks.

Recent work has begun to address this gap: PEAR evaluates security 
vulnerabilities in planner-executor multi-agent 
architectures~\citep{dong2026pear}, and TAMAS benchmarks adversarial 
risks across six attack vectors specific to multi-agent interactions 
including impersonation and colluding 
agents~\citep{tamas}. Nevertheless, coverage remains far 
from comprehensive. Multi-agent security testing should additionally 
evaluate the abilities of multiple agents to work together to overcome 
safeguards even when a single agent cannot~\citep{Jones2024}; the 
robustness of cooperation between networked agents in the presence of 
malicious adversaries~\citep{Barbi2025}, including the effects of network 
topology and interaction protocol~\citep{Huang2024,Marro2024,Hammond2025}; 
the ability of agents to adversarially manipulate or extract information 
from other agents or humans, especially in 
tandem~\citep{wu_inference_2024,wei_trustworthy_2024}; and security 
vulnerabilities specifically designed to propagate through inter-agent 
interactions~\citep{Gu2024,Lee2024,Ju2024}. Adversarial testing, 
including leveraging advanced AI 
adversaries~\citep{perez2022red,pavlova2024automatedredteaminggoat}, 
should also be applied to non-AI entities that AI agents will soon be 
able to interact with. For more complex entities or larger networks of 
agents, more tractable anticipatory modelling tools such as 
ABMs~\citep{vestad_survey_2024} may also be necessary.

Addressing these limitations requires rethinking how security benchmarks 
for multi-agent systems are designed and evaluated. Benchmarks should 
move toward explicit agent modeling, instantiating multiple agents with 
distinct roles, partial observability - i.e., no single agent has 
access to the complete system state or all relevant information needed 
to solve the task alone - and limited communication, such that 
coordination is necessary rather than optional. They should model 
co-evolving attacker and defender teams, enabling evaluation of strategic 
adaptation rather than static task completion. Evaluation metrics must 
go beyond task success to capture coordination efficiency, communication 
quality, robustness to miscoordination, and safety compliance. 
Underpinning all of this is the need for long-horizon, stateful 
environments with memory and delayed consequences, without which 
realistic cyber operations cannot be faithfully represented.

\subsection{Sociotechnical Security Defences}
As with many of the risks presented in this report, security risks are inherently sociotechnical in nature and can therefore benefit from improved AI governance as well as technical solutions.
For example, regulators could codify security standards for multi-agent systems in safety-critical domains and assign responsibility to organizations deploying unsecure multi-agent systems so as to ensure sufficient investment in security~\citep{khlaaf2023}.
Tools such as software bills of materials~\citep{sbom} and lineage tracking~\citep{lineagetracking} can bolster transparency in this regard.
Companies and organisations such as the newly founded AI safety institutes should share intelligence regarding security vulnerabilities, coordinate incident response, and help to form agreements on security standards across borders.
More generally, we must work to ensure that different stakeholders possess an appropriate degree of transparency, participation, and accountability in navigating difficult trade-offs between the security, performance, and privacy of interactions between advanced AI agents~\citep{sangwan_cybersecurity_2023,Gabriel2024}.
This work would benefit greatly from collaboration with security experts and distributed systems engineers as well as social scientists and policymakers. A fundamentally important mitigation strategy against social engineering attacks is to strengthen human users through education~\citep{montanez_human_2020}.

In addition, mitigations must address the broader sociotechnical threat surface identified in the Background and Taxonomy section. Human operators require structured training in multi-agent failure modes, adversarial manipulation, and secure configuration practices, supported by regular red-team exercises and incident response simulations~\citep{nist_ai_rmf_2023,mitre_attck,nist_sp_800_61r3}. Access to effective supervision mechanisms is equally critical: operators should be equipped with interpretable monitoring tools, comprehensive logging, anomaly detection systems, and clearly defined escalation pathways~\citep{human_factors_cybersecurity_2025}. Role-based access controls, separation of duties, and zero-trust principles can further mitigate insider risks and unauthorized lateral movement between agents~\citep{wylde_zero_2021,greitzer2010insider}. Finally, organizations should foster a strong security culture - encouraging responsible disclosure, protecting whistleblowers, and learning from near-misses - to enhance collective resilience against sociotechnical threats~\citep{dekker2012just,nist_ai_rmf_2023}.


\section{Multi-Agent Security Governance}
\label{sec:multi-agent_ai_governance}




The challenges cataloged in Sections~\ref{sec:taxonomy} and~\ref{sec:challenges} do not admit purely technical solutions. Even with successful deployment of the infrastructure, detection mechanisms, and protocols discussed in Section~\ref{sec:challenges}, multi-agent systems raise inherently normative questions regarding accountability, responsibility, and authority. This section discusses three such challenges, namely that: compositionality breaks the unit of regulatory analysis, emergence breaks the chain of liability, and collective-action dynamics make infrastructure adoption itself a governance challenge.

\subsection{Why Multi-Agent Governance is Different}
\label{ss:why_different}

Current AI governance focuses on single entities, whether a model, a developer, or a deployer, as the object of regulation. Certification attaches to the model; liability flows to the deployer; audit traces a single chain of responsibility. Two properties of multi-agent systems complicate this focus.

\emph{Compositionality.} Individually safe components compose into unsafe systems~\cite{Ghosh2025securityframework, spera2026safetynoncompositionalformalframework}. \cite{Jones2024} demonstrate that an adversary can combine a refusal-trained model, used for benign-but-hard subtasks, with a weaker but more permissive model, used for malicious-but-easy steps, to achieve harm that neither component could produce alone. \cite{brundage2026frontieraiauditingrigorous} explicitly flag multi-agent composition as an open auditing problem with no current methodology.

\emph{Emergence.} The threats catalogued in Section~\ref{sec:taxonomy}, including covert collusion, cascades, mixed-motive coordination failures, and heterogeneous attacks, do not stem from a single adversarial agent, instead arising from interaction dynamics between agents. For example, \cite{syrnikov2026institutionalaigoverningllm} show that LLM agents individually constrained by prompt-level safety instructions converge on collusive pricing in Cournot simulations, and \cite{deshpande2026strategicaicournotmarkets} find that sustained tacit coordination can drive prices well above competitive equilibrium without explicit communication. In each case the emergent outcome violates the intent of every individual constraint in place.

\subsection{Accountability and Liability}
\label{ss:accountability}

Harms arising from the interaction of agents belonging to different principals pose challenges for attributing liability. Cascade failures of the kind discussed in \S\ref{cascade_att}, where localized compromise propagates through a network of well-behaved agents, have no proximate cause in the sense tort law requires. Emergent collusion of the kind demonstrated by \cite{syrnikov2026institutionalaigoverningllm} and \cite{motwani_secret_2024} has no attributable intent, which also challenges doctrines built around mens rea. Heterogeneous attacks of the kind demonstrated by \cite{Jones2024} may involve no component operating outside specification, which product-liability claims anchored to component defect are thus ill-suited to address.

Single-agent AI liability is already unresolved, with responsibility distributed ambiguously among model providers, deployers, and end users. Multi-agent deployment compounds this in three ways. Firstly, multiple principals participate, each with partial control and partial visibility. The technical threat-attribution problem discussed in \S\ref{threat_attr} is often unsolvable under worst-case assumptions, leaving no defensible causal chain on which to attach responsibility. And harms may be time-distributed or statistical, visible only at the population level, in which case no individual instance supports a claim.








\subsection{Collective Action and Coordination}
\label{ss:collective_action}

Additionally, achieving the coordination required to implement technical solutions to multi-agent security issues is itself a governance challenge. The technical infrastructure proposed in Section~\ref{sec:challenges}, including identity registries, logging systems, permission manifests, and protocol-level governance primitives, must be adopted across a heterogeneous ecosystem of actors with divergent interests. Three dynamics distinguish this from ordinary infrastructure governance.

\emph{Tragedy of the commons.} Ecosystem-wide security can be modeled as a public good. An actor who invests in robust monitoring, restrictive protocols, or audit infrastructure bears the full cost but captures only a fraction of the benefit, most of which accrues to other participants. Standard public-goods logic predicts systematic under-investment. The \cite{syrnikov2026institutionalaigoverningllm} results illustrate the same pattern at the behavioural level: actors given declarative constraints without enforceable consequences do not internalize the social costs of their coordination choices.

\emph{Weakest-link dynamics.} Multi-agent security properties may be bounded by the least-secure component in the system. A single MCP server with poisoned tool definitions compromises every agent that connects to it~\cite{hou2025model, beber2026escaperoute}. A single deployer running unaudited agents contaminates shared environments. A single defecting principal creates an opening for others to follow. Under weakest-link dynamics, compliance by major actors is insufficient, which inverts the standard regulatory strategy of concentrating enforcement on high-leverage firms and implies that voluntary frameworks such as the IMDA guidance discussed below are unlikely to produce adequate protection.

\emph{Information and action-space asymmetries.} No single actor possesses the visibility required to design effective system-level defences. Model providers see their models but not downstream deployments. Deployers see their own compositions but not others operating in shared environments. Regulators see neither at the granularity required. Even with universal willingness to contribute, the knowledge and authority needed to design coordinated defences is distributed in a way that no unilateral or bilateral arrangement can aggregate.

These dynamics make the infrastructure proposals of \cite{chan2025infrastructure} and \cite{marro2026permissionmanifestswebagents} harder to realize in practice, independent of their technical merit. Identity registries, logging infrastructure, and protocol-level hooks must be adopted collectively to work, and governed once adopted. Further questions arise regarding which actors are responsible for building, operating, and controlling this substrate. Fully decentralized arrangements face the familiar bootstrapping problem: without coordinating authority, early adoption is unattractive, and without early adoption the network effects that justify the infrastructure never materialize. Centralized arrangements create chokepoints, since whoever operates the identity registry or the dominant agent protocol may control market access for a substantial portion of the ecosystem.

\subsection{The Limits of Model-Level Governance}
\label{ss:limits_of_model_level}

Current governance proposals are insufficient for addressing these challenges.

\emph{Institutional frameworks.} The EU AI Act (Regulation 2024/1689)~\cite{euaiact2024}, the most developed AI regulatory instrument globally, does not mention multi-agent risks; its general-purpose AI provisions (Articles~51--56) target the foundation-model layer and inherit the model-centric unit of analysis~\cite{oueslati2025ahead}. In the United States, NIST's COSAiS project~\cite{nist_cosais_2025} is developing control overlays that include multi-agent use cases, though finalized controls remain in development; the CAISI AI Agent Standards Initiative~\cite{nist_caisi_2026} has made agent authentication and identity infrastructure dedicated research pillars; and the NIST NCCoE concept paper on AI Agent Identity and Authorization~\cite{nist_caisi_rfi_fr_2026} addresses authentication and least-privilege access but remains at consultation. Singapore's IMDA Model AI Governance Framework for Agentic AI~\cite{imda2026agentic}, published in January 2026, is the first governance framework explicitly designed for agentic systems and the only one that acknowledges multi-agent coordination risks, including cascading errors and unintended coordination. The OECD Expert Group on Agentic AI~\cite{oecd2026agentic} and the International AI Safety Report~\cite{bengio2026international} establish conceptual groundwork but propose no binding mechanisms.

\emph{Interaction protocols.} Two complementary protocols have emerged as de facto standards for agent communication: the Model Context Protocol (MCP), introduced by Anthropic in November 2024 and subsequently donated to the Agentic AI Foundation under the Linux Foundation in December 2025, standardizing agent-to-tool communication via JSON-RPC~2.0; and the Agent-to-Agent (A2A) protocol, announced by Google in April 2025 and contributed to the Linux Foundation in June 2025, enabling agent discovery and delegated execution (both discussed in Section~\ref{sec:background}). Neither was designed primarily for security, and both exhibit significant vulnerabilities.
 
MCP's security surface is well-documented~\citep{hou2025mcp}. Beyond 
the tool poisoning and rug-pull attacks discussed in 
Section~\ref{sec:tool-use}, MCP introduces several additional vectors. 
Sandbox escape attacks enable agents to break out of their execution 
environment, as demonstrated by CVE-2025-53109 and CVE-2025-53110 
targeting Anthropic's filesystem MCP 
server~\citep{beber2026escaperoute}. MCP's sampling feature, which 
allows servers to request LLM completions from the client, introduces 
further risks including resource theft and covert tool 
invocation~\citep{huang2025mcp}. The June 2025 MCP specification 
mandated OAuth~2.1 for authentication, but adoption remains 
inconsistent across implementations.
 
Similarly, A2A's Agent Cards (JSON metadata files hosted at \texttt{/.well-known/agent.json}) enable capability discovery, authentication requirements, and trust establishment. However, because Agent Cards are processed as trusted data by consuming agents, they constitute an attractive surface for prompt injection. A2A shares MCP's exposure to name-spoofing and rug-pull attacks. The Ripple Effect Protocol \cite{chopraRippleEffectProtocol2025} proposes a complementary coordination mechanism in which agents share lightweight sensitivity information to accelerate alignment, but securing such open coordination channels against the collusion and manipulation threats catalogued in Section~\ref{sec:taxonomy} remains an open problem.
 
Critically, neither MCP nor A2A was designed to facilitate \emph{secure} coordination between agents. Both standardize message transport and capability discovery; neither specifies governance primitives such as enforceable action constraints, audit-grade provenance, or mechanisms for detecting the collusion patterns analyzed in \S\ref{collusion_exfil}. The governance layer that \cite{chan2025infrastructure} argue is necessary for multi-agent ecosystems does not yet exist at the protocol level.

\emph{Runtime governance mechanisms} The collusion results of \cite{syrnikov2026institutionalaigoverningllm}, discussed in \S\ref{ss:why_different}, provide the strongest empirical case for runtime enforcement over declarative constraints. Their governance graphs are public, immutable manifests that specify legal states, allowed transitions, and sanctions for violations; an Oracle/Controller runtime interprets the manifest and maintains a cryptographically keyed, append-only governance log. The decisive finding is not merely that governance graphs reduce collusion, but that prompt-level ``constitutional'' constraints produce no measurable effect, demonstrating that declarative prohibitions do not bind under optimization pressure and that runtime enforcement with consequences is necessary.

Effective governance requires evaluation methodologies adequate to the risks. No evaluation framework addresses system-level risks analogous to financial systemic risk: correlated failures across agent populations sharing common model providers, concentration risk from dependence on shared infrastructure (e.g., a single MCP server serving multiple agents), or cascading disruption through shared environments. The 2010 Flash Crash (Case Study~3.5) and the collusion results of \cite{syrnikov2026institutionalaigoverningllm} and \cite{deshpande2026strategicaicournotmarkets} illustrate that such systemic effects are not hypothetical. \cite{reuel2025openproblemstechnicalai} note that evaluations can establish only lower bounds on capabilities, a limitation dramatically amplified in multi-agent compositions where emergent properties exceed the sum of individually measured components.

\pagebreak
\section*{Conclusion}

The emergence of distributed or decentralized ecosystems populated by autonomous, goal‐driven AI agents has exposed a rich terrain of security challenges that lie beyond the traditional boundaries of cybersecurity and AI safety. In this work, we have argued for the establishment of \emph{multi‐agent security} as a distinct field dedicated to understanding and mitigating worst‐case threats in systems of interacting AI. By surveying a broad taxonomy of vulnerabilities - from covert steganographic collusion and adversarial stealth to cascade dynamics at the edge of chaos - we have highlighted how adaptive communication protocols, emergent behavior, and multipolar attributions together conspire to undermine conventional defenses.  

Crucially, the open problems in multi‐agent security are not merely technical curiosities but constitute \textbf{fundamental barriers to the safe deployment of next‐generation AI infrastructures}. Issues such as robust threat attribution in diffuse networks, the detection of secret collusion channels, and the characterization of systemic instabilities resist reduction to isolated solution recipes. Instead, they demand a concerted research agenda that embraces the interplay between dynamic agent behaviors, adversarial incentives, and the evolving structure of decentralized platforms.  

By drawing attention to these uncharted challenges - rather than prescribing narrow mitigation strategies - our aim is to catalyze a community‐wide effort to develop principled frameworks, analytical tools, and evaluation methodologies tailored to multi‐agent contexts. Only through such collective exploration can we hope to unveil the theoretical limits of cooperative and adversarial interactions, identify the boundaries of safe operating regimes, and chart a path toward resilient, accountable, and transparent multi‐agent ecosystems.  

\section*{Acknowledgements}

The authors express their sincere gratitude for the generous support provided by the Royal Academy of Engineering (UK), OpenAI, the Foresight Institute, Schmidt Sciences, the Cooperative AI Foundation, and numerous other contributors. CSDW also thanks Prof. Philip H.S. Torr (Torr Vision Group, University of Oxford) for his ongoing support.


\section*{Epilogue: Security at the ``Edge of Chaos'' - A Long-Term Vision}
\label{sec:edge}

\textit{This section paints a tentative future vision for what security could mean in the era of decentralized superintelligence.}

Theories of collective intelligence posit that emergent capabilities arise when systems operate at the so‑called \emph{edge of chaos}, a critical regime balancing order and randomness~\citep{Langton1990,Kauffman1993} that generate \textit{complex} phenomenology~\citep{krakauerLargeLanguageModels2025} that is technically different from merely \textit{complicated} dynamics~\citep{shapiraAgentsChaos2026}. Indeed, the ``patchwork AGI'' hypothesis suggests that general capability levels may first manifest not through monolithic systems but through coordination among sub-AGI agents with complementary skills and affordances~\citep{tomasev_distributional_2025}. This regime yields maximal adaptability and creativity but also introduces profound security challenges. First, the inherent unpredictability and nonlinear state transitions at the edge of chaos hinder traditional verification and static analysis techniques, leaving vulnerabilities that adversaries can exploit~\citep{Newman2018}. Second, the rapid propagation of perturbations characteristic of critical networks can amplify localized attacks into global disruptions, akin to epidemic cascades in scale‑free graphs~\citep{Pastor-Satorras2001,Buldyrev2010}. Third, defensive interventions that disregard the system's critical balance may themselves trigger adverse emergent behaviors, effectively pushing the network into chaotic or overly rigid regimes~\citep{Kauffman1993}. Finally, securing such systems demands runtime, adaptive defenses that detect anomalies in evolving interaction patterns rather than relying on fixed signatures, and that embed self‑healing mechanisms inspired by biological robustness~\citep{Kitano2004}. Together, these strategies form the foundation of a security‑by‑design approach tailored to the edge-of‑chaos regime in decentralized AI.






\pagebreak

\bibliography{main}
\bibliographystyle{tmlr}

\appendix

\end{document}